\documentclass[10pt,superscriptaddress,nofootinbib,twocolumn]{revtex4-2}
\usepackage{amsthm,amsmath,amssymb}
\usepackage{mathrsfs}
\usepackage{amsthm}
\usepackage{amsmath}
\usepackage{amssymb}
\usepackage{graphicx}
\usepackage{epstopdf}
\usepackage{url}
\usepackage{float}
\usepackage{bm}
\usepackage{ifthen}
\usepackage[usenames,dvipsnames]{color}
\usepackage{mathrsfs}
\usepackage[colorlinks=true,citecolor=blue,urlcolor=black]{hyperref}
\usepackage{float}
\usepackage{subfigure}
\usepackage{multirow}
\usepackage{makecell}

\newcommand{\be}{\begin{equation}}
\newcommand{\ee}{\end{equation}}

\newcommand\xrowht[2][0]{\addstackgap[.5\dimexpr#2\relax]{\vphantom{#1}}}
\newcommand{\sket}[1]{{\ensuremath{\lvert#1\rangle}}}
\newcommand{\lket}[1]{{\ensuremath{\left\lvert#1\right\rangle}}}
\newcommand{\ket}[1]{\if@display\lket{#1}\else\sket{#1}\fi}

\usepackage{stackengine}
\newcommand{\sbra}[1]{{\ensuremath{\langle#1\rvert}}}
\newcommand{\lbra}[1]{{\ensuremath{\left\langle#1\right\rvert}}}
\newcommand{\bra}[1]{\if@display\lbra{#1}\else\sbra{#1}\fi}

\newcommand{\sbraket}[2]{{\ensuremath{\langle#1\rvert#2\rangle}}}
\newcommand{\lbraket}[2]{{\ensuremath{\left\langle#1\!\left\rvert\vphantom{#1}#2\right.\!\right\rangle}}}
\newcommand{\braket}[2]{\if@display\lbraket{#1}{#2}\else\sbraket{#1}{#2}\fi}

\newcommand{\sketbra}[2]{{\ensuremath{\lvert #1\rangle\!\langle #2\rvert}}}
\newcommand{\lketbra}[2]{{\ensuremath{\left\lvert #1\right\rangle\!\!\left\langle #2\right\rvert}}}
\newcommand{\ketbra}[2]{\if@display\lketbra{#1}{#2}\else\sketbra{#1}{#2}\fi}

\theoremstyle{plain}

\theoremstyle{definition}

\begin{document}

\title{Breaking the Rate-Loss Bound of Quantum Key Distribution with Asynchronous Two-Photon Interference}
\author{Yuan-Mei Xie}\thanks{These authors contributed equally to this work}	
\author{Yu-Shuo Lu}\thanks{These authors contributed equally to this work}
\author{Chen-Xun Weng}
\author{Xiao-Yu Cao}
\author{Zhao-Ying Jia}
\author{Yu Bao}
\author{Yang Wang}
\affiliation{National Laboratory of Solid State Microstructures, School of Physics and Collaborative Innovation Center of Advanced Microstructures, Nanjing University, Nanjing 210093, China.}
\author{Yao Fu}
\affiliation{MatricTime Digital Technology Co. Ltd., Nanjing 211899, China}	
\author{Hua-Lei Yin}\email{hlyin@nju.edu.cn}
\affiliation{National Laboratory of Solid State Microstructures, School of Physics and Collaborative Innovation Center of Advanced Microstructures, Nanjing University, Nanjing 210093, China.}
\author{Zeng-Bing Chen}\email{zbchen@nju.edu.cn}
\affiliation{National Laboratory of Solid State Microstructures, School of Physics and Collaborative Innovation Center of Advanced Microstructures, Nanjing University, Nanjing 210093, China.}
\affiliation{MatricTime Digital Technology Co. Ltd., Nanjing 211899, China}		
%%%%%%%%%%%%%%%%%%%%%%%%%%%%%%%%%%%%%%%%%

\begin{abstract}
Twin-field quantum key distribution can overcome the  {secret key capacity of repeaterless quantum key distribution} via single-photon interference.	 However, to compensate for the channel fluctuations and lock the laser fluctuations, the  techniques of phase tracking and phase locking are indispensable in experiment, which drastically increase experimental complexity and hinder free-space realization.
We herein present an asynchronous measurement-device-independent quantum key distribution protocol that can surpass the  {secret key capacity} even without phase tracking and phase locking. Leveraging the concept of time multiplexing, asynchronous two-photon Bell-state measurement is realized by postmatching two interference detection events. For a 1 GHz system, the new protocol reaches a transmission distance of 450 km without phase tracking. After further removing phase locking, our protocol is still capable of breaking  {the capacity at 270 km}. Intriguingly, when using the same experimental techniques, our protocol has a higher key rate than the phase-matching-type twin-field protocol. In the presence of imperfect intensity modulation, it also has a significant advantage in terms of the transmission distance over the sending-or-not-sending type twin-field protocol. With high key rates and accessible technology,  our work provides a promising candidate for practical scalable quantum communication networks.
\end{abstract}

\maketitle
\section{Introduction}
Quantum key distribution (QKD)~\cite{bennett2014quantum, ekert1991quantum} allows the distribution of information-theoretically secure keys guaranteed by quantum mechanical limits. However, experimental implementations of QKD always deviate from the theoretical assumptions used in security proofs, leading to various quantum hacking attacks~\cite{zhao2008quantum,lydersen2010hacking,tang2013source,xu2020secure,pirandola2020advances}. 
Fortunately, all security loopholes on the detection side are closed by measurement-device-independent QKD (MDIQKD)~\cite{lo2012measurement}, which introduces an untrusted third party, Charlie, to perform two-photon Bell-state measurement in the intermediate node. Thus far, MDIQKD has made many theoretical and experimental breakthroughs~\cite{braunstein2012side,Liu2013Experimental,rubenok2013real,zhou2016making, experiment2014mdi,fu2015long,curty2014finite,yin2014long,comandar2016quantum,yin2016measurement,semenenko2020chip,yin2017experimental,wei2020high,woodward2021gigahertz,wang17measurement,zheng2021heterogeneously, tang2016measurement}.

However, because a significant number of photons are inevitably lost in the channel, the key rate of most QKD protocols, including MDIQKD, is rigorously limited by the  {secret key capacity of repeaterless QKD~\cite{pirandola2009direct,takeoka2014fundamental, pirandola2017fundamental,pirandola2019end, das2021universal}}, more precisely, the Pirandola--Laurenza--Ottaviani--Banchi (PLOB) bound $R = -\log_2 (1-\eta)$~\cite{pirandola2017fundamental}, where $R$ is the secret key rate and $\eta$ is the total channel transmittance between the two users. Utilizing single-photon interference, twin-field QKD (TFQKD)~\cite{lucamarini2018overcoming} and its variants~\cite{ma2018phase,wang2018twin,yin2019measurement,lin2018simple,curty2019simple,cui2019twin,yin2019coherent,hu2019sending,zeng2020symmetry,xie2021overcoming,wang2020optimized,li2021long},  such as sending-or-not-sending QKD~(SNSQKD)~\cite{wang2018twin} and phase-matching QKD~(PMQKD)~\cite{ma2018phase,zeng2020symmetry},  have been proposed to increase the key rate to $O(\sqrt{\eta})$, overcoming the PLOB bound. Since then, they have aroused widespread concern.  For example, remarkable progress has been made in the theory of finite key analysis~\cite{maeda2019repeaterless, yin2019finite, jiang2019unconditional,curras2021tight}.
Additionally, some notable experimental implementations	have been reported~\cite{minder2019experimental,zhong2019proof,wang2019beating,liu2019experimental,fang2020implementation,chen2020sending,liu2021field,zhong2021proof,chen2021twin,clivati2022coherent,Pittaluga2021600-km,chen2021quantum,wang2022twin}. The longest transmission distance of more than  {830 km} was recently achieved in the laboratory through optical fibers  {~\cite{wang2022twin}.}	

Because the phase evolution of the twin fields is sensitive to both channel length drift and frequency difference between two user lasers, phase-tracking and phase-locking techniques are vital for twin-field-type protocols. Phase tracking is used to compensate for the phase  {fluctuation on the channels connecting the users to Charlie}, where bright reference light pulses are sent to measure the phase fluctuation. However, the performance of the system is severely affected because the bright light causes scattering noise and occupies the time of the quantum signal~\cite{minder2019experimental,wang2019beating,liu2019experimental,fang2020implementation,chen2020sending,liu2021field,chen2021twin}. Phase tracking also imposes a high counting requirement on the detectors. 	 {Phase locking is utilized to lock the frequency and the phase of the two users’ lasers. There are several types of phase-locking techniques, including laser injection~\cite{comandar2016quantum}, optical phase-locked loop~\cite{kazovsky1986balanced, kazovsky19901320}, and  time-frequency dissemination technology~\cite{liu2019experimental,Predehl441}.} However, they all require additional channels between users to transfer the reference light. In addition, laser injection may introduce security risks~\cite{huang2019laser,pang2020hacking}, and the optical phase-locked loop and the time-frequency dissemination  technology  both require complicated feedback systems. An ingenious replacement for phase locking and phase tracking in TFQKD experiments~\cite{zhong2019proof,zhong2021proof} is the plug-and-play type construction~\cite{yin2019measurement}, but it is susceptible to Trojan horse attacks~\cite{Lucamarini2015practical,Sajeed2016insecurity,Zhang2021Securing,bozzio2021multiphoton}.  Furthermore, free-space realization of QKD in various types of channels, including the atmosphere~\cite{cao2020long,liu2020optical}, seawater~\cite{hu2021decoy}, and satellite-to-ground channels~\cite{chen2021integrated,dequal2021feasibility,yin2020entanglement,takenaka2017satellite}, is essential to establishing a global-scale quantum networks~\cite{wehner2018quantum, sidhu2021advance}.  {However, deploying phase-tracking and phase-locking techniques in free space faces some  technical challenges. For example, phase locking requires additional channels between the two users. In summary, these technical requirements increase experimental complexity, may incur security risks, and hinder the implementation of TFQKD in commerce and free space.}

In this work, we propose an asynchronous-MDIQKD protocol to remove these requirements. It has a simple hardware implementation while enjoying a high key rate.  Recall that in the conventional time-bin encoding MDIQKD scheme~\cite{ma2012alternative}, coincidence detection of two neighboring time bins is required, resulting in the  $O({\eta})$ scaling of the key rate~\cite{das2021universal}. Intriguingly, we observe that the two neighboring time bins can actually be decoupled. Specifically, the requirement for coincidence detection of two neighboring time bins is unnecessary. By utilizing time multiplexing, we match two detected time bins that are phase-correlated to establish an asynchronous two-photon Bell state, and the key rate is enhanced to $O(\sqrt{\eta})$. This can be regarded as breaking the recently proposed linear boundary of dual-rail protocols~\cite{das2021universal}. We highlight the intrinsic differences between  time-bin encoding MDIQKD and polarization encoding  MDIQKD---time multiplexing is possible solely for time-bin encoding, where there are infinite time modes. For polarization encoding, only two orthogonal modes exist.

Because the differential phase evolution of each time bin is approximately equal in a short time interval, we can postmatch two phase-related time bins without the phase-tracking and phase-locking techniques at the cost of a slight increase in the interference error rate. We show that after removing these techniques, the  misalignment angle  between the two users is approximately $0.1 \pi$ within 1 $\mu $s with simple commercially available instruments,  giving rise to an interference error rate of approximately $2.4 \% $.  In this case, our protocol can beat the PLOB bound at a distance of approximately 270 km, achieving a reasonable trade-off between practicality and performance.  { By circumventing the need for phase locking and phase tracking, our protocol can directly utilize techniques derived from free-space MDIQKD~\cite{cao2020long}, thereby facilitating QKD to break the secret key capacity in free space}. Moreover, when the imperfection in light intensity modulation is considered ~\cite{xie2021scalable}, our protocol achieves a longer transmission distance and  higher key rates compared with SNSQKD with actively odd-parity pairing (AOPP) ~\cite{xu2020sending,jiang2020zigzag} and PMQKD~\cite{zeng2020symmetry} when using the same experimental techniques. 

\begin{figure}[t!]
\centering
\includegraphics[width=8.6cm]{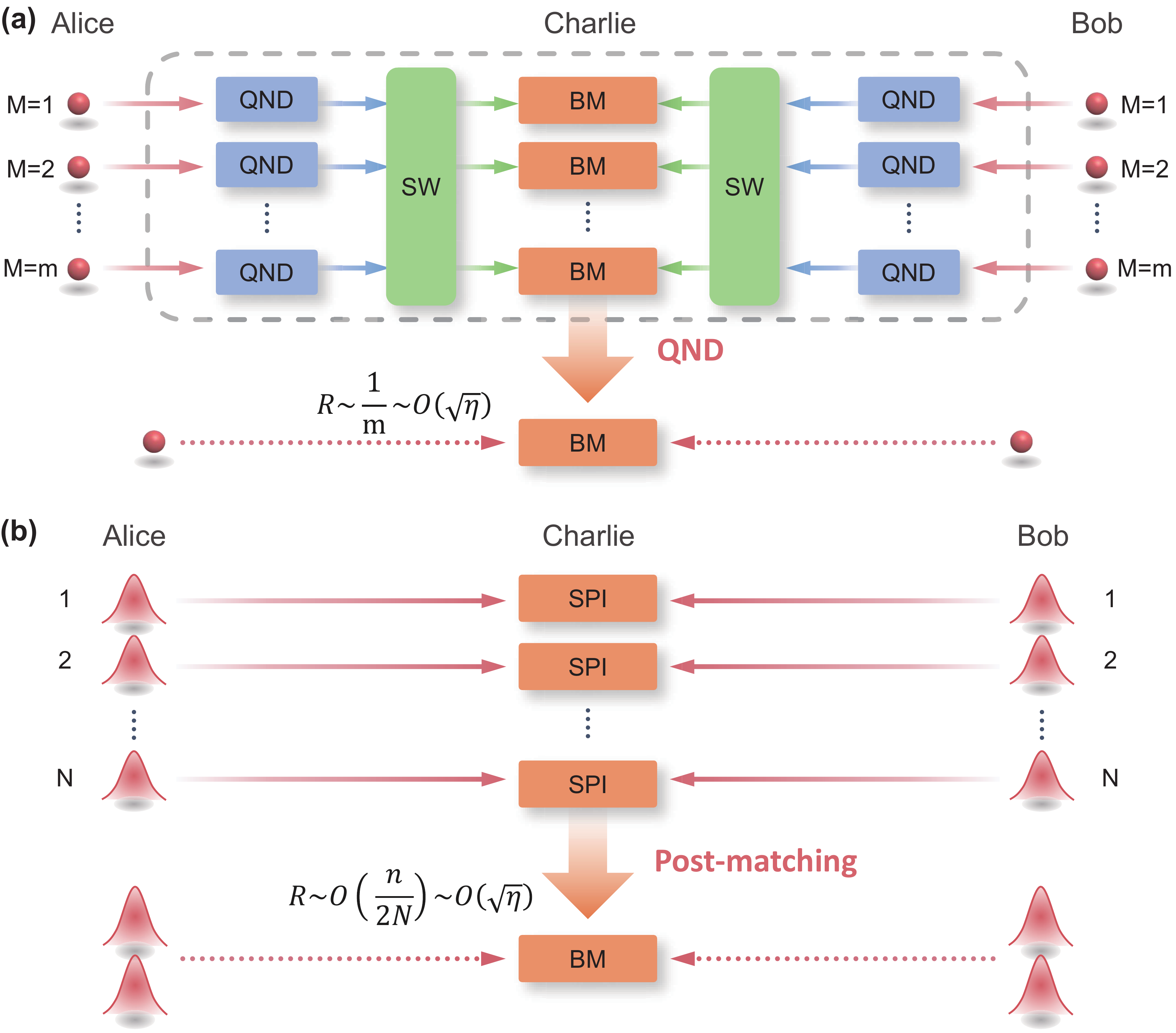}
\caption{Basic idea of the adaptive MDIQKD scheme~\cite{azuma2015all} and this work. (a) In the adaptive MDIQKD scheme, Alice and Bob send $m$ single-photon pulses to Charlie using spatial multiplexing.  Charlie applies quantum nondemolition measurement (QND) to confirm the arrival of pulses, matches the arriving pulses using optical switches (SW), and performs two-photon Bell measurements (BM) on the pairs.   (b) In our asynchronous-MDIQKD protocol, Alice and Bob send $N$ pulses to Charlie to perform single-photon interference (SPI), and a two-photon Bell state is obtained by postmatching two successful SPI events. } \label{fig_basic_idea} 
\end{figure}

\section{asynchronous-MDIQKD protocol}\label{sec_p1}
In this section, we first introduce the basic idea of the asynchronous-MDIQKD protocol and then present a detailed protocol description.

\subsection{Protocol topology}
Our proposal is motivated by the  adaptive MDIQKD scheme~\cite{azuma2015all}, as shown in Fig.~\ref{fig_basic_idea}(a). In this scheme, the idea of spatial multiplexing is leveraged, where $m$ optical pulses in single-photon states pass  through $m$ channels. When $m\sqrt{\eta}\ge 1$, one or more single photons arrive at Charlie from Alice and Bob each, with unitary probability, resulting in $O(\sqrt{\eta})$ scaling with the key rate. 

In the time-bin encoding MDIQKD~\cite{ma2012alternative}, the information is encoded in the relative phase between the optical modes in two separate time bins $i$, $j$, with $i=2t-1$, $j=2t$, where $t$ is an integer. Let $\ket{1,0}^{ij}=\ket{1}^{i}\ket{0}^{j}$ denote the quantum state, where there is one photon in time bin $i$ and zero photon in time bin $j$.  Alice and Bob prepare quantum states $\ket{+z}=\ket{1,0}^{i,j}$, $\ket{-z} =\ket{0,1}^{i,j}$, and $\ket{\pm x}=(\ket{1,0}^{i,j}\pm\ket{0,1}^{i,j})/\sqrt{2}$, and send them to Charlie for Bell-state measurement, where $\ket{\pm z}$  are the eigenstates in the $Z$ basis,  and $\ket{\pm x}$ are the eigenstates in the $X$ basis. Note that for the Bell states announced by Charlie, $\ket{\psi^\pm}=1/\sqrt{2}(\ket{1,0}^{i,j}_a\ket{0,1}^{i,j}_b\pm\ket{0,1}^{i,j}_a\ket{1,0}^{i,j}_b)$, where subscript $a$ represents mode ``Alice'' and $b$ represents mode ``Bob'',  we can decouple $i$ and $j$ as two independent variables~\cite{bose2013duality},  making time multiplexing possible.

The basic idea of our asynchronous-MDIQKD protocol is illustrated in Fig.~\ref{fig_basic_idea}(b). $N$ pairs of optical pulses are sent in $N$ time bins. In each time bin, single-photon interference is performed, and $n\approx N\sqrt{\eta}$ successful detection events are obtained.  By time multiplexing, Alice and Bob can postmatch successful detection events of two time bins that are phase-correlated to establish two-photon entangled states $\ket{\psi^{\pm}}$, leading to the $O(\sqrt{\eta})$ decay of the key rate.

\begin{figure}[t]
\centering
\includegraphics[width=8.6cm]{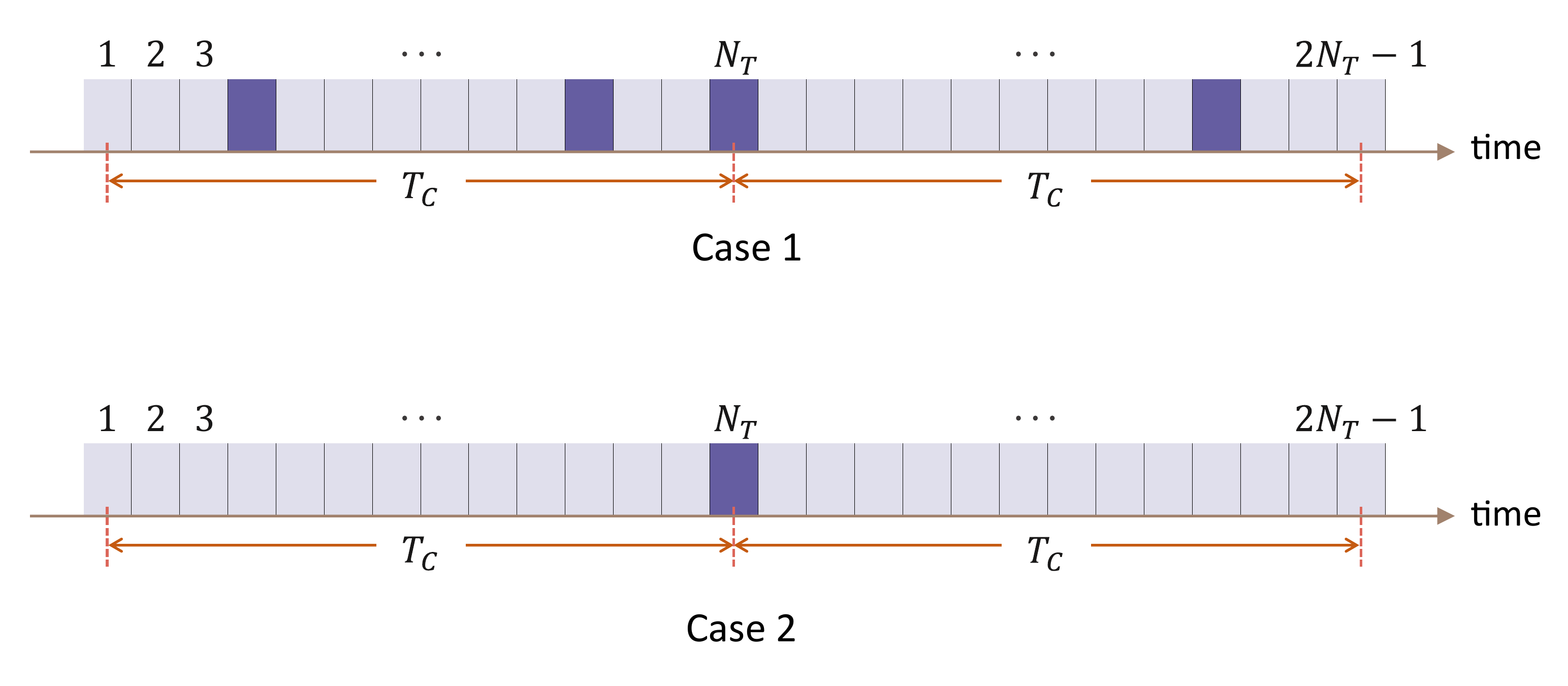}
\caption{Schematic of $\mathit{case~1}$ and $\mathit{case~2}$. Each time bin is represented by a lattice, and detection events are painted in deep purple.
	$\mathit{case~1}$: a detection event around which other detection events can be found within $T_c$; 	$\mathit{case~2}$: a detection event around which there are no other detection events within $T_c$. $N_T$ is the total number of pulses sent within $T_c$, which is proportional to the system repetition rate. 
}\label{fig_case1_2}
\end{figure}

We consider the case in which Alice and Bob always match two successful detection events within a short time interval $T_c$, where $T_c$ is on the order of microseconds. Hereafter, we abbreviate a successful detection event to a detection event for simplicity. As shown in Fig.~\ref{fig_case1_2},  the detection events can be classified  into two cases: $\mathit{case~1}$ and $\mathit{case~2}$.  The $\mathit{case~1}$ event indicates that other detection events can be found around it within $T_c$, and the differential phase evolution of detection events in $T_c$ are unknown but almost the same. The $\mathit{case~2}$ event indicates that there are no other detection events around it within $T_c$. After removing phase tracking and phase locking, the differential phase evolution of each $\mathit{case~2}$ event becomes indeterminate. This means that $\mathit{case~2}$ events have no phase correlations with each other. These events should be carefully handled to ensure security. For example, Charlie can know the total photon number in each optical pulse pair sent by Alice and Bob via quantum nondemolition measurements, and Charlie always lets the joint single-photon be detected as $\mathit{case~2}$ events. Charlie's operation  would not be found by Alice and Bob. If Alice and Bob discard $\mathit{case~2}$ events, the single-photon pairs cannot be reasonably estimated using the conventional decoy-state method. Fortunately, a sufficiently small number (one on average) of $\mathit{case~2}$ events will not affect the security of the protocol.

\begin{figure}[t]
\centering
\includegraphics[width=8.6cm]{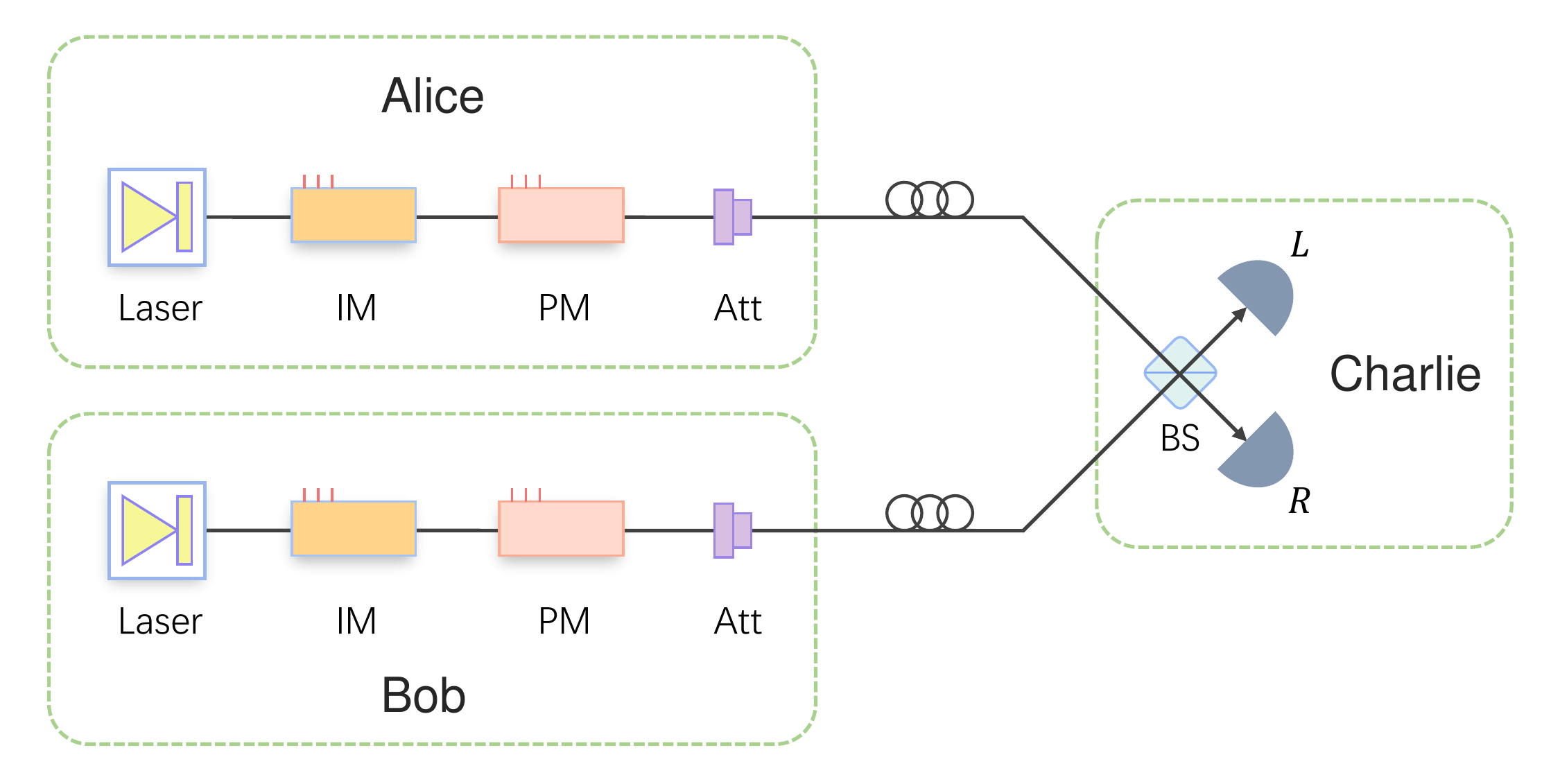}
\caption{Schematic of the setup for the asynchronous-MDIQKD protocol. Alice and Bob utilize a narrow-linewidth continuous-wave laser, intensity modulator (IM), phase modulator (PM), and attenuator (ATT) to prepare phase-randomized weak coherent pulses with different intensities and phases. Charlie performs interference measurement with a beam splitter (BS) and  single-photon detectors. Charlie announces the detection events where only the detector $\mathbf{L}$ or $\mathbf{R}$ clicks.	}\label{figuresetup}
\end{figure}	

\subsection{Protocol description}
The schematic of the  asynchronous-MDIQKD setup is shown in Fig.~\ref{figuresetup}, and the details of the protocol are presented as follows.

{\it{1.~Preparation.}} Alice and Bob repeat the first two steps for $N$ rounds to obtain sufficient data. At each time bin $i\in\{1,2,\ldots,N\}$, both phase $\theta_{a}^{i}$ $\in[0,2\pi)$ and  classical bit $r_{a}^{i}\in\{0, 1\}$ are randomly chosen by Alice. Alice then prepares a weak coherent pulse $\ket{e^{\textbf{i}(\theta_{a}^{i}+r_{a}^{i}\pi)}\sqrt{k_{a}^i}}$ with probability $p_{k_a}$,  where $k_a^i\in \{\mu_a,~\nu_a,~\mathbf{o}_{a},~\hat{\mathbf{o}}_{a}\}$ corresponds to the signal, decoy, preserve-vacuum, and declare-vacuum intensities, respectively ($\mu_a>\nu_a>\mathbf{o}_{a} 
=\hat{\mathbf{o}}_{a} = 0$). 
Similarly, Bob prepares a phase-randomized weak coherent pulse. Alice and Bob send the corresponding pulses $\ket{e^{\textbf{i}(\theta_{a}^{i}+r_{a}^{i}\pi)}\sqrt{k_{a}^i}}$ and $\ket{e^{\textbf{i}(\theta_{b}^{i}+r_{b}^{i}\pi)}\sqrt{k_{b}^i}}$ ($k_b^i\in \{\mu_b,~\nu_b,~\mathbf{o}_{b},~\hat{\mathbf{o}}_{b}\}$) to Charlie via insecure quantum channels.

{\it{2.~Measurement.}} For each time bin $i$, Charlie performs interference measurement on the two received pulses.  Charlie obtains a detection event when only one detector clicks. He publicly announces whether a detection event is obtained and which detector clicked. 
In the following description, we define $\{k_a,~k_b\}$ as a detection event when Alice sends intensity $k_a$, and Bob sends $k_b$. The compressed notation $\{k_a^ik_a^{j},~ k_b^ik_b^{j}\}$ indicates that$\{k_a^i,~k_b^i\}$ and $\{k_a^j,~ k_b^j\}$ are matched, the first label referring to time bin $i$, and the second to time bin $j$.

{\it{3.~Sifting.}} Alice and Bob first check the number of $\mathit{case~2}$ events. If the number of occurrences of $\mathit{case~2}$ is smaller than or equal to $\Lambda$, the data of $\mathit{case~2}$ can be discarded, where $\Lambda$ is a preset threshold; otherwise, they abort the protocol. For $\mathit{case~1}$ events, when at least either Alice or Bob chooses  a decoy or  declare-vacuum intensity,
they announce their intensities and phase information through authenticated channels. They then use the following rules to randomly match two $\mathit{case~1}$ events with a time interval of less than $T_c$.

The unannounced detection events $\{\mu_a,~\mathbf{o}_{b}\}$, $\{\mu_a,~\mu_b\}$, $\{\mathbf{o}_{a},~\mu_b\}$, and $\{\mathbf{o}_{a},~\mathbf{o}_{b}\}$ are used to  form data in the $Z$  basis. For these events, Alice randomly matches a time bin $i$ of intensity  $\mu_a$  with another time bin $j$ of intensity $\mathbf{o}_{a}$. Alice and Bob discard detection events that cannot find a matchable peer. Then, Alice sets her bit value  to 0 (1) if 
$i< j~(i>j)$ and informs Bob of the serial numbers $i$ and $j$. In the corresponding time bins, if Bob chooses intensities $k_b^{\min\{i,j\}}=\mu_b$~$(\mathbf{o}_{b})$ and $k_b^{\max\{i,j\}}=\mathbf{o}_{b}$~$(\mu_b)$, the bit value is set to 0 (1). Bob announces an event where $k_b^i$ $=$ $k_b^j=\mathbf{o}_{b}$ or $\mu_b$. Thus, the valid events in the $Z$ basis are $\{\mu_a \mathbf{o}_{a},~ \mathbf{o}_{b}\mu_b\}$, $\{\mu_a\mathbf{o}_{a},~\mu_b\mathbf{o}_{b}\}$, $\{\mathbf{o}_{a}\mu_a,~ \mathbf{o}_{b}\mu_b\}$, and $\{\mathbf{o}_{a}\mu_a,~\mu_b\mathbf{o}_{b}\}$.

The detection events $\{\nu_a,~\nu_b\}$, $\{o_a,~\nu_b\}$, $\{\nu_a,~o_b\}$, $\{\hat{\mathbf{o}}_a,o_b\}$, and $\{\mathbf{o}_a,~\hat{\mathbf{o}}_b\}$  are used to form  data in the $X$ basis, where $o_{a(b)}\in\{\mathbf{o}_{a(b)},~\hat{\mathbf{o}}_{a(b)}\}$. The  global phase of Alice (Bob) at time bin $i$ is defined as  $\varphi^i_{a(b)}:= \theta^i_{a(b)} + \phi^i_{a(b)}$, where $\phi^i_{a(b)} $ is the phase evolution from the channel.  The global phase difference between Alice and Bob at time bin $i$ is $\varphi^i =\varphi_a^i - \varphi_b^i$. Alice and Bob randomly choose two detection events that satisfy $k_{a}^i=k_{a}^j$, $k_{b}^i=k_{b}^j$ and $\left|\varphi^i -\varphi^j\right|= 0 $ or $\pi$
(experimental techniques for quantifying $\left|\varphi^i -\varphi^j\right|$ are discussed later in Sec.~\ref{sec_discussion}). They then match the two events as $\{k_a^ik_a^j,~k_b^ik_b^j\}$. By calculating the classical bits $r_a^i \oplus r_a^j $ and $r_b^i \oplus r_b^j $, Alice and Bob obtain a bit value in the $X$ basis, respectively.	Afterwards, in the $Z$ basis, Bob always flips his bit. 	In the $X$ basis, Bob flips part of his bits to correctly correlate them with Alice’s (see Table.~\ref{taba}).

{\it{4.~Parameter estimation.}} Alice and Bob exploit the random bits from the  $Z$ basis to form the $n^z$-length raw key bit. The remaining bits in the  $Z$ basis are used to calculate the bit error rate  $E^z$. They reveal all bit values in the $X$ basis to obtain the total number of errors. The decoy-state method~\cite{lo2005decoy,wang2005beating} is utilized to estimate the number of vacuum events in the $Z$ basis $s_{0\mu_b}^{z}$, number of single-photon pairs  $s_{11}^z$, bit error rate in the $X$ basis $e_{11}^x$, and phase error rate of single-photon pairs $\phi_{11}^z$ in the $Z$ basis~(see Appendix~\ref{app_simulation} for details).

\emph{Note}: To estimate the single-photon component gain of each postmatching interval, we assume that the single-photon distributions in all detection events are independent and identical.

{\it{5.~Postprocessing.}} Alice and Bob distill the final keys by using the error correction algorithm with $\varepsilon_{\rm{cor}}$-correct, and the privacy amplification algorithm with $\varepsilon_{\rm{sec}}$ -secret. Similar to Ref.~\cite{curty2014finite},  the length of the final secret key $\ell$  with total security $\varepsilon_{\rm{AMDI}}=\varepsilon_{\rm{sec}} + \varepsilon_{\rm{cor}}$ can be given by
\begin{equation}
\begin{aligned}\label{eq_keyrate_finite}
	\ell=&\underline{s}_{0\mu_b}^z+\underline{s}_{11}^z\left[1-H_2(\overline{\phi}_{11}^z)\right]-\lambda_{\rm{EC}} \\
	&-\log_2\frac{2}{\varepsilon_{\rm cor}}-2\log_2\frac{2}{\varepsilon'\hat{\varepsilon}}-2\log_2\frac{1}{2\varepsilon_{\rm PA}},
\end{aligned}
\end{equation}
where $\underline{x}$ and $\overline{x}$ denote the lower and upper bounds of the observed value $x$, respectively.  $\lambda_{\rm{EC}}=n^zfH_2(E^z)$ is the amount of information leaked during error correction, where  $f$ is the error correction efficiency,  and $H_2(x)=-x\log_2x-(1-x)\log_2(1-x)$ is the binary Shannon entropy function. $\varepsilon_{\rm cor}$ is the failure probability of error verification, and  $\varepsilon_{\rm PA}$ refers to the failure probability of privacy amplification. $\varepsilon'$ and $\hat{\varepsilon}$ represent the coefficients when using the chain rules of smooth min-entropy and max-entropy, respectively. $\varepsilon_{\rm sec}=2(\varepsilon'+\hat{\varepsilon}+2\varepsilon_e)+\varepsilon_\beta+\varepsilon_0+\varepsilon_1+\varepsilon_{\rm PA}$, where $\varepsilon_0$, $\varepsilon_1$ and $\varepsilon_e$ are the failure probabilities of estimating the terms $s_{0\mu_b}^z$, $s_{11}^z$, and $\phi_{11}^z $, respectively.

\begin{table}[b]
\centering
\caption{Postprocessing of bit in the sifting step.  In the $X$ basis, Bob decides whether to implement a bit flip to guarantee correct correlations, depending on the clicking detectors announced by Charles and the global phase difference between two matching time bins. Here, $\mathbf{RL}$~($\mathbf{LR}$) denotes the detectors $\mathbf{R}$~($\mathbf{L}$) and $\mathbf{L}$~($\mathbf{R}$) clicks at time bins $i$ and $j$, respectively. $\mathbf{RR}$~($\mathbf{LL}$) denotes that the detector $\mathbf{R}$~($\mathbf{L}$) clicks at time bins $i$ and $j$.}\label{taba}
\begin{tabular}[b]{ccccc}
	\hline
	\hline
	~ &\multicolumn{2}{c}{Measurement results of Charlie}~  \\
	\hline
	Global phase difference~ ~& ~~~~$\mathbf{RL}$ ~($\mathbf{LR}$) ~~ & $\mathbf{RR}$~($\mathbf{LL}$)~~   \\
	\hline
	~~~  $|\varphi^i -\varphi^j| = 0$  ~~&  ~~~~~~Bit flip~~~~~ & No bit flip  ~~   \\
	\hline
	~~~  $|\varphi^i -\varphi^j| =\pi$ ~~&~~~~No bit flip~~~ & Bit flip ~~  \\
	\hline
\end{tabular}
\end{table}

\section{Experimental Discussion}
\label{sec_discussion}

In the asynchronous-MDIQKD protocol, the information in the interference mode is encoded in the phase difference of two matched time bins. One may regard the latter as the reference mode and the former as the signal mode, which is the same as the time-bin encoding MDIQKD~\cite{ma2012alternative}. In the $X$ basis, Alice and Bob postmatch pulses of two time bins $i$ and $j$ with phase relation  $\left|\varphi^i -\varphi^j\right|= 0 $ or $\pi$, where $\varphi^{i(j)} = \theta_a^{i(j)}-\theta_b^{i(j)}+\phi^{i(j)}$ is the global phase difference at time bin $i~(j)$. $\theta_a^{i(j)}$ and $\theta_b^{i(j)}$ are random phases known to Alice and Bob.  $\phi^{i(j)} = \phi^{i(j)}_a - \phi^{i(j)}_b$ is the differential phase evolution at time bin $i~(j)$, which is determined by the frequency difference between the two users' lasers and the fluctuation of the fiber channels.
When the two pulses sent by Alice and Bob reach Charlie, the phase evolutions of the two pulses are $\phi_a^i= 2\pi v_a^i (t^i - l^i_a/s)$ and $\phi_b^i = 2\pi v_b^i (t^i - l^i_b/s)$, respectively, where $t^i$ is the time of the time bin $i$, $v_{a(b)}^i$ is the laser frequency, $s$ is the speed of light in the fiber, and $l_{a(b)}^i$ is the fiber length between Alice (Bob) and Charlie at time bin $i$. In symmetric channels, the differential phase evolution between Alice and Bob can be expressed as
\begin{equation}
\begin{aligned}
	\phi^i=\phi_a^i -\phi_b^i 
	=2\pi \delta v^it^i -  \frac{2\pi}{s}( \delta v^i l^i +v^i \delta l^i 
	),
\end{aligned}
\end{equation}
where $\delta v^i =v_a^i - v_b^i$, $\delta l^i = l_a^i - l_b^i$, $l^i = (l^i_a+ l^i_b)/2 $ and  $v^i =(v^i_a+v^i_b)/2$. Ensuring the phase correlation between time bins $i$ and $j$ is essential for postmatching. In the experiment, when the time interval for postmatching $T_c$ is large, the phase correlation can be maintained by using  phase-tracking and phase-locking techniques to measure differential phase evolution in each time bin. Fortunately, when $T_c$ is small, the phase correlation naturally exists, that is, the differential phase evolution is approximately constant within $T_c$ because the fiber length drift rate and relative phase drift rate between lasers are relatively small. Therefore, our protocol can discard phase-tracking and phase-locking techniques at the cost of a slight increase in the interference error rate.

\subsection{Removing phase tracking}\label{simplified technique}
Assuming that the frequencies of the two user lasers are synchronized with phase-locking techniques, $v_a^{i(j)} = v_b^{i(j)} =v^{i(j)} $, we have 
\begin{equation}
\begin{aligned}
	\label{eq_phi_ab_ij}
	\phi^j -\phi^i =\frac{2\pi}{s}
	\left(v^j \delta l^j - v^i \delta l^i\right).
\end{aligned}
\end{equation}
When $T_c$ is on the order of tens of microseconds, say 50 $\mu $s, the frequency drift is small for typical commercially available narrow-linewidth lasers. Hence, $\phi^j -\phi^i$ is mainly determined by the relative phase drift caused by the fiber length drift, which was measured to be approximately $8$ rad/ms $@$ $402$ km~\cite{fang2020implementation}, corresponding to a phase drift of 0.4 rad per $ 50~\rm{\mu s}$. Note that two time bins $i$ and $j$ are randomly and uniformly distributed within the time interval $T_c$; therefore, the mean phase drift between the two time bins is half of the maximum value. Consequently, there will be an  intrinsic  interference error rate of approximately $(1-\cos{0.2})/2\approx 1 \%$.

\begin{figure}[t]
\centering
\includegraphics[width=8.6cm]{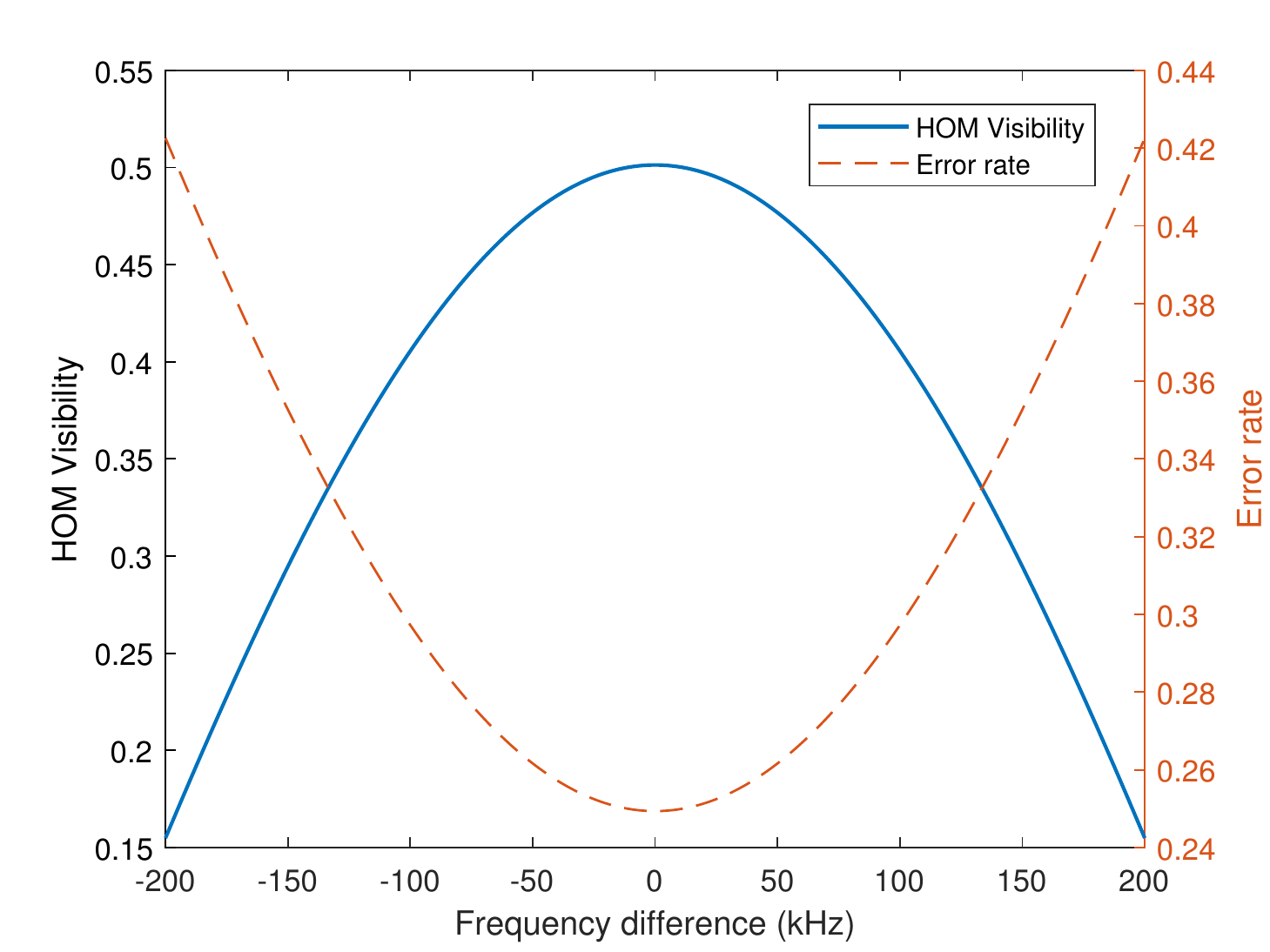}
	\caption{HOM visibility and error rate as a function of frequency difference between two lasers $\delta v$. We set the time interval of two consecutive time bins $\tau=1~\mu$s. The blue line is the HOM visibility, and the orange dotted  line is the interference error rate.}\label{fig_HOM}
\end{figure}

This indicates that our protocol with  short-term matching can be experimentally implemented without  phase tracking when the matching interval $T_c$ is on the order of tens of microseconds. Note that sufficient detection counts should be accumulated per $T_c$ for postmatching. The detection count per $T_c$ can be approximated as $ T_cF(1-e^{-\overline{\mu} \eta_{d}\sqrt{\eta_{\rm{ch}}} })$, where  $F$ is the system frequency, $\eta_{\rm{ch}}$ is the channel transmittance between Alice and Bob, $\eta_{d}$ is the detection efficiency, and  $\overline{\mu}$ is the total mean photon number of Alice and Bob. At 400 km, by using a 1 GHz system with $\eta_{d} = 70 \%$ and ultra-low loss fiber,  there will be approximately 9.3 detection events per $T_c$ if we set $\bar{\mu}=0.5$, which is sufficient for postmatching.

\subsection{Removing  phase tracking and phase locking}
We consider the case in which neither phase-tracking nor phase-locking techniques are applied.	When $T_c$ is on the  order of a few tens of microseconds, the fiber length drift is also negligible. The relative phase drift between time bins $i$ and $j$ is mainly determined by the frequency difference between the two independent lasers, which can be expressed as
\begin{equation} 
\begin{aligned}
	\label{eq_phi_ab_ij_no_PhaseLocking}
	\phi^j -\phi^i =2 \pi \delta v (t^j-t^i).
\end{aligned}
\end{equation}
In practice, $\delta v$ is the sum of stable laser frequency difference between the two users and laser frequency random drift. The former can be a relatively large value (up to tens of megahertz) known to Alice and Bob. The latter is an unknown small value (tens of kilohertz). For simplicity, below we discuss the case where the two users try to adjust their lasers to nearly the same frequency, leaving only an unknown small $\delta v $. In this case, there is an intrinsic phase misalignment $\pi \delta v T_c$ on average during $T_c$. If $\delta v$ is controlled to 100~kHz, for $T_c =$ 1 $\rm{\mu s}$, $\phi^i -\phi^j \approx 0.1 \pi $, resulting in an interference error rate of 2.4$\%$.  By using an experimental setup with $F = 10$ GHz~(which is feasible under current technology~\cite{takesue2007quantum, islam2017provably}) and $\eta_{d} = 70\%$, at $300$ km, there will be approximately $11.7$ detection events per microsecond if we set $\overline{\mu}=0.5$.  

\begin{figure}[t]
\centering
\includegraphics[width=8.6cm]{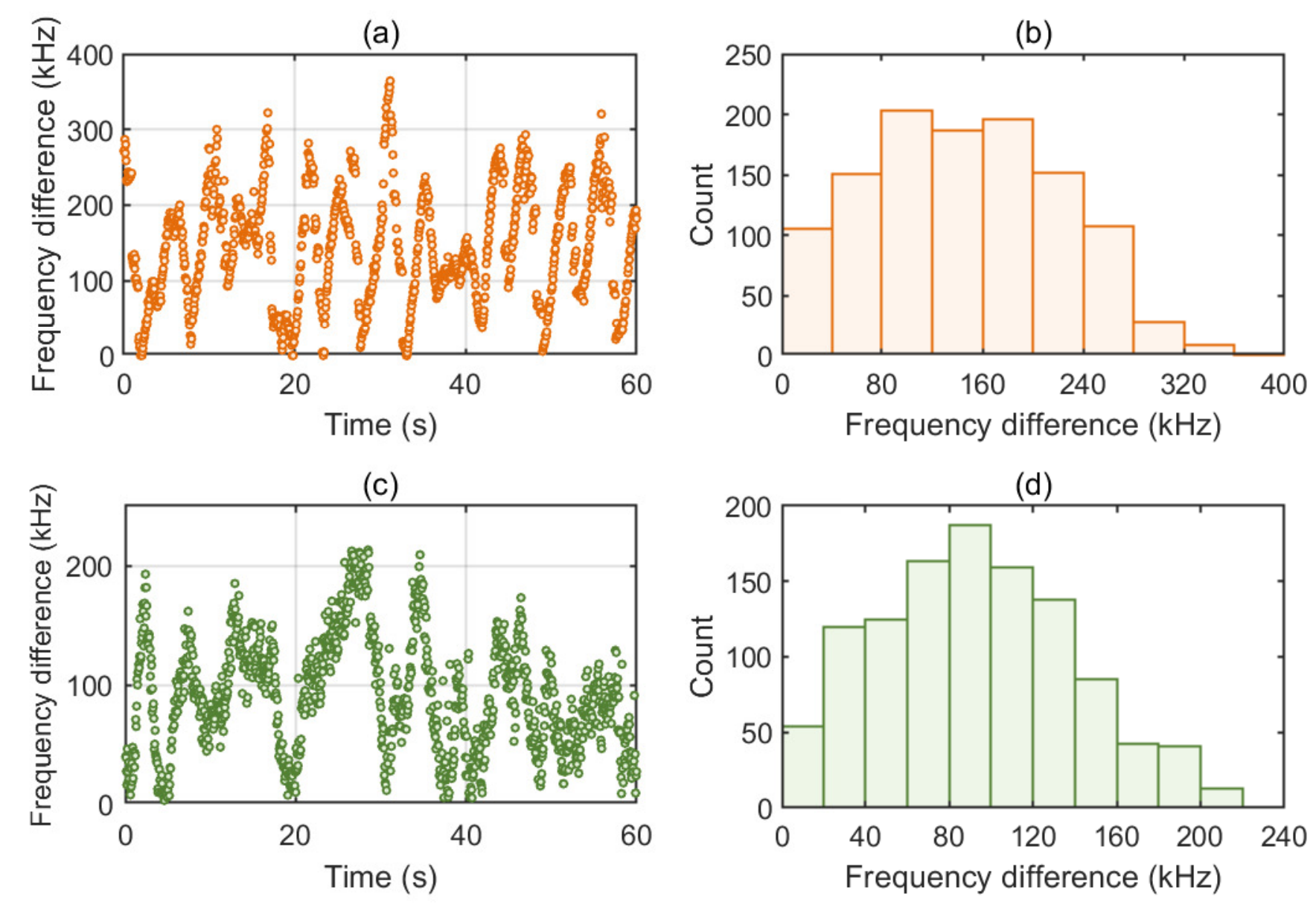}
\caption{Test results of the frequency difference between two independent lasers , which are  obtained by manually adjusting the temperature. (a) Frequency difference between two NKT lasers for 60 s. (b) Histogram of the frequency difference distribution between the two NKT lasers. The mean frequency difference is  145 kHz. (c) Frequency difference between two RIO Orion lasers of 60 s. (d) Histogram of the frequency difference distribution between two RIO Orion lasers, where the mean value is  92.5 kHz.  Note that if using automatic feedback systems, the frequency difference can be further reduced.}\label{fig_experiment}
\end{figure}

The experimental requirement of the asynchronous-MDIQKD protocol without phase tracking and phase locking is similar to that of the previous phase encoding MDIQKD, in which the information is encoded in the relative phase of the two time bins with time interval $\tau$. The frequency difference between the two users will inevitably misalign the phase basis, leading to intrinsic phase misalignment $\delta \phi =  2\pi \delta v \tau $~\cite{woodward2021gigahertz}. In experimental demonstrations in  Refs.~\cite{yin2016measurement} and~\cite{woodward2021gigahertz}, the time delays are 6.37 ns and 0.5 ns, respectively. The maximum frequency difference is 37.5 MHz and 30 MHz, respectively, which introduce  phase misalignments of 0.47 $\pi$ and 0.03 $\pi$, respectively. For our  protocol with short time matching, $\tau$ is on the order of microseconds, and the frequency difference is approximately tens of kilohertz.         

In a real setup, to suppress the frequency difference between two independent lasers within $ 100$ kHz, Alice and Bob can periodically enter the calibration process to calibrate the frequency of their lasers. There are several approaches available for frequency calibration. The most straightforward way is to measure the beat note of two lasers.  Alice and Bob can also locally calibrate the frequency with a frequency standard~\cite{cao2020long}. In addition,
they can measure the Hong--Ou--Mandel~(HOM) interference and utilize the interference visibility as the feedback signal~\cite{tang2016measurement} to minimize the frequency difference. These are mature techniques in time-bin MDIQKD.  

We briefly describe the method of frequency calibration using HOM interference~\cite{tang2016measurement} as follows. Consider the HOM interference of two weak coherent pulses with the same time, polarization and spatial mode, but different frequencies. The HOM interference visibility is related to the time interval of two consecutive time bins $\tau$  and the frequency difference  $\delta v $. As shown in Fig.~\ref{fig_HOM}, we simulate the interference visibility $V$ and interference error rate $E=(1- V)/2$ as a function of $\delta v $, where $\tau = 1~\mu$s. The visibility reaches the maximum value of approximately 0.5 when the frequency difference is 0, and decreases rapidly as the frequency difference increases. The minimum error rate of approximately  $25\%$ is obtained when $\delta v = 0$. When  $\delta v$ increases to 100 kHz, the error rate increases to 29.7$\%$. In experiment, one can adjust the laser frequency to minimize the error rate. When the observed error rate is close to 25$\%$, the frequencies of the two users’ lasers are calibrated. By increasing $\tau$, the error rate becomes more sensitive to $\delta v$.

\begin{figure}[t]
\centering
\includegraphics[width=8.6cm]{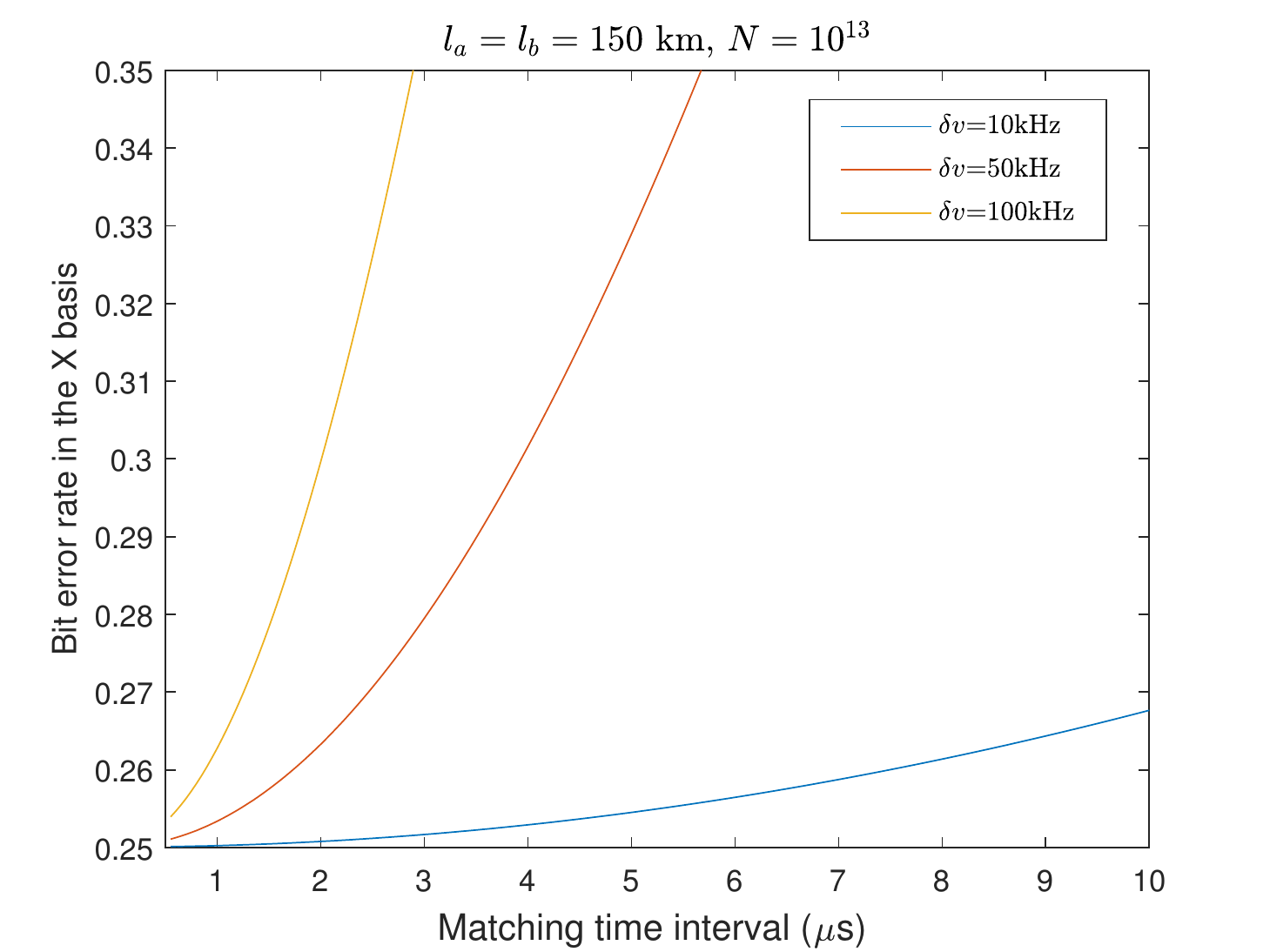}
	\caption{ {The bit error rate in the $X$ basis as a function of the matching time interval $T_c$, where both phase tracking and phase locking are not adopted.  The distance between Alice and Bob is $300$ km and the   system repetition rate is $F=4$ GHz. The frequency difference $\delta v$ is set to 10, 50 and 100 kHz. }}\label{fig_error_noFreqLock}
\end{figure}

To demonstrate the feasibility of the experimental setup without phase-tracking and phase-locking techniques, we measured the frequency difference between two independent lasers. Two narrow-linewidth lasers working at 1550.12 nm emit continuous light.  The continuous light passes through fiber, interferes at a beam splitter, and is detected by a photoelectric detector. The beat note was recorded using an oscilloscope. Fig.~\ref{fig_experiment} (a) shows the beat frequencies of the two NKT lasers (Koheras BASIK E15). These lasers support fine piezoelectric tuning with a minimum tuning frequency on the order of kHz.	During the test time of 60 s, we manually adjusted the frequency of one of the lasers every few seconds to minimize the observed beat frequency. A histogram of the recorded data is presented in Fig.~\ref{fig_experiment}(b), and the mean value is  145 kHz.  We also measured the  frequency difference between the two independent Rio lasers (Rio ORION). They were first tuned to the same frequency and kept free running during the 60 s test time. The collected data are shown in Fig.~\ref{fig_experiment}(c), and the histogram is shown in Fig.~\ref{fig_experiment}(d), with an average of  92.5 kHz. In a real setup, if using the RIO laser or the NKT laser, the frequency needs to be adjusted every few seconds. We stress that the frequency difference can be further reduced by utilizing automatic feedback systems and improving the stability of the experimental environment. 

Additionally, we simulate the bit error rate in the X basis as a function of the time interval $T_c$  in Fig.~\ref{fig_error_noFreqLock}, which is calculated under the optimal key rate. The distance between Alice and Bob is 300 km and the system repetition rate is $F= 4$ GHz. The experimental parameters were set to the typical values given in Table.~\ref{tab1}. With a fixed frequency difference, the error rate in the X basis increases as the time bin interval increases. When the frequency difference $\delta v=10$ kHz, the error rate increases slowly, indicating that a relatively low system repetition rate is sufficient for the experiment.

\begin{figure}[t]
\centering
\includegraphics[width=8.6cm]{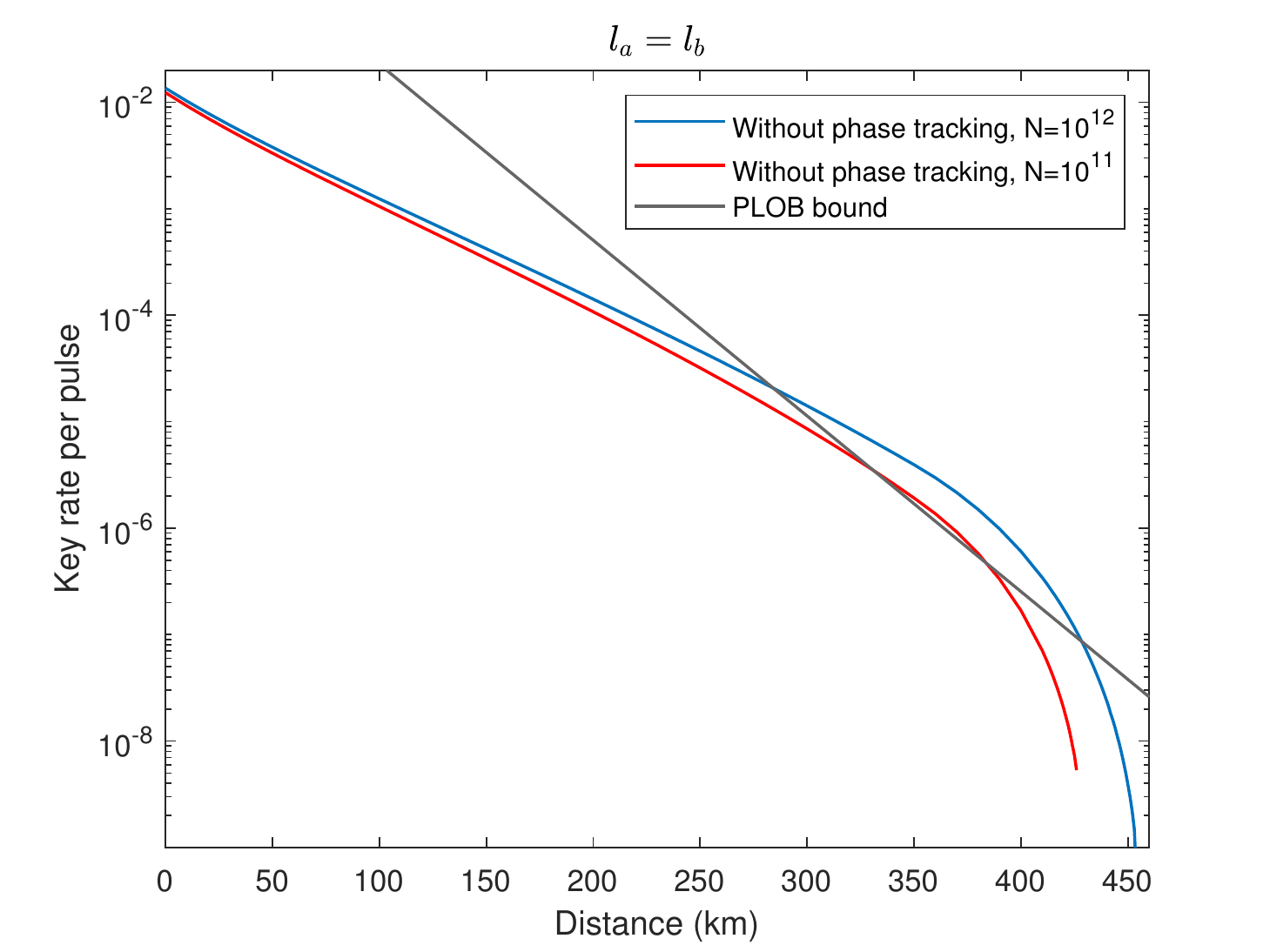}
\caption{Secret key rates of the asynchronous-MDIQKD with short time matching as a function of the distance  when implemented without phase tracking.  {Here, we set the matching time interval $T_c=50~\mu $s, the system repetition rate is $F= 1$ GHz and the angle of misalignment in the $X$ basis $\sigma =\pi/10$.}  Our protocol can break the PLOB bound at a distance of approximately 280 km with $N=10^{12}$, and the transmission distance reaches 450 km.}\label{fig_keyrate_noStrongRef}
\end{figure}

\begin{table}[b!]
\centering
\caption{Simulation parameters. $\eta_{d}$ and $p_{d}$ are the detector efficiency and dark count rate, respectively.   $\alpha$ is the attenuation coefficient of the fiber and $f$ denotes the error correction efficiency. $\epsilon$ is the failure probability considered in the error verification and finite data analysis.}\label{tab1}
\begin{tabular}[b]{@{\extracolsep{10pt}}ccccc}
	\hline
	\hline
	$\eta_{d}$  & $p_{d}$ & $\alpha$ & $f$  & $\epsilon$\\
	\hline\xrowht{7pt}
	$70\%$ & $10^{-8}$   & $0.165$ dB/km & $1.1$  &
	$36/23\times10^{-10}$\\
	\hline
	\hline
\end{tabular}
\end{table}

\section{Performance and discussion}
\label{sec_performance_APMDI}
We numerically simulate the key rate $R=\ell/N$ of our asynchronous-MDIQKD protocol in finite-size cases. We set the threshold $\Lambda=10$. The genetic algorithm is exploited to globally search for the optimal value of light intensities and their corresponding probabilities. When optimizing the key rate, we set an additional condition in which the mean number of $\mathit{case~2}$ events  $\bar{N}^{c_2}\le1$. Assuming that the distribution of the number of events in $\mathit{case~2}$ follows a Poisson distribution, the probability that the observed number of $\mathit{case~2}$ events in the experiment exceeds $\Lambda$ is  $1-\sum_{k=0}^{10}\frac{\varsigma^j}{j!}e^{-\varsigma}\approx1\times10^{-8}$, where $\varsigma=\bar{N}^{c_2}$. This reveals the robustness of our protocol, which will only fail once in 100 million rounds of experiments.

The experimental parameters were set to the typical values given in Table.~\ref{tab1}.  We set the failure parameters $\varepsilon'$, $\hat{\varepsilon}$, $\varepsilon_e$, $\varepsilon_\beta$, and $\varepsilon_{\rm PA}$ to be the same $\epsilon$.  We denote the distance between Alice (Bob) and Charlie as $l_a$ ($l_b$). In the symmetric case,  that is, $l_a = l_b = l/2$, we have $\varepsilon_0+\varepsilon_1=14\epsilon$ because  the Chernoff bound~\cite{chernoff1952measure, yin2020tight} is used 14 times to estimate $s_{0\mu_b}^z$, $s_{11}^z$, and $e_{11}^x$. The corresponding security bound is $\varepsilon_{\rm{asyn}}=2.4\times 10^{-9}$. Similarly, in the asymmetric case, we have $\varepsilon_0+\varepsilon_1=13\epsilon$ and $\varepsilon_{\rm{asyn}}=2.3\times 10^{-9}$.

\begin{figure}[t]
\centering
\includegraphics[width=8.6cm]{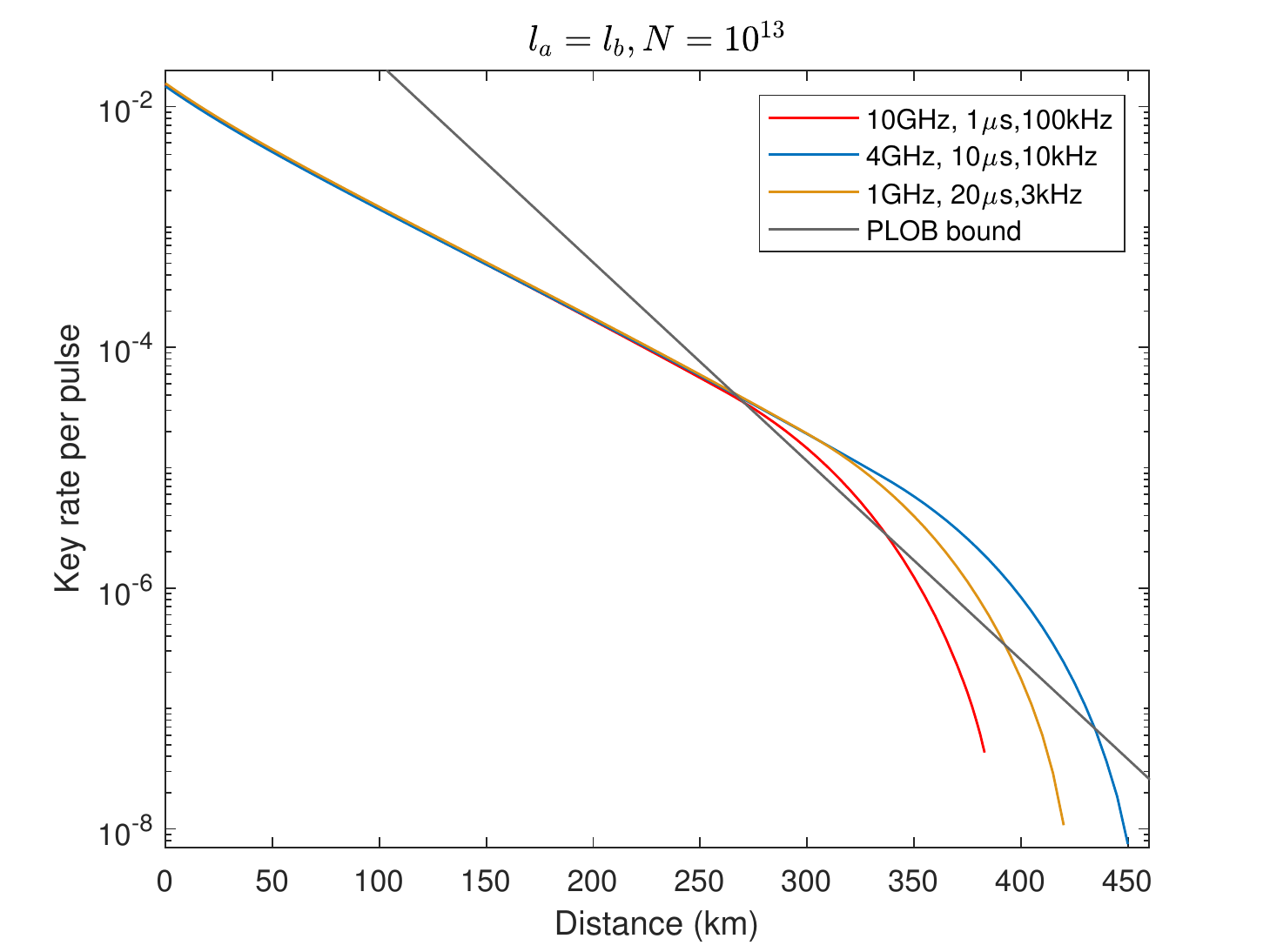}
\caption{ Secret key rates of our protocol with short-term matching  as a function of the distance where both phase tracking and phase locking are not adopted. Different system repetition rates $F$, matching time intervals $T_c$ and frequency differences $\delta v$ are taken into consideration. Red line: $F=10$ GHz, $T_c=1~\mu $s, $\delta v=100$ kHz; blue line:  $F=4$ GHz, $T_c=10 ~\mu $s, $\delta v=10$ kHz; yellow line: $F=1$ GHz, $T_c=20 ~\mu $s, $\delta v=3$ kHz. Our  protocol  can overcome the PLOB bound at 270 km with a key rate of $2 \times 10^{-5}$.}\label{fig_keyrate_noFreqLock}
\end{figure}

First, we calculate the key rates of the asynchronous-MDIQKD protocol with a short time interval $T_c$.  The detailed formulas for simulating our protocol are presented in Appendix~\ref{simu_pb}. The statistical fluctuation analysis formulas are presented in Appendix~\ref{statistical}. Fig.~\ref{fig_keyrate_noStrongRef} shows a scenario in which phase tracking is removed. The time interval $T_c = 50$ $\rm{\mu s}$, and the system repetition rate is $F = 1$ GHz. We assume that the angle of misalignment in the $X$ basis $\sigma=\pi/10$. The key rate beats the PLOB bound at 280 km under the condition where the data size is $N = 10^{12}$ and the transmission distance reaches 450 km.  One can also transmit over more than 420 km and overcome the PLOB even with a data size of $N=10^{11}$.

The key rate in the case where neither phase-tracking nor phase-locking techniques are employed is shown in Fig.~\ref{fig_keyrate_noFreqLock}.  {Here, we consider the frequency differences of 3, 10 and 100~kHz. Correspondingly, the system repetition rates are 1 GHz, 4 GHz (which has been employed in the experiments in Ref.~\cite{wang2022twin}) and 10 GHz. Note that when $T_c = 20~\mu $s, the phase misalignment caused by fiber phase drift is considered. The simulation results show that the proposed protocol can overcome the PLOB bound at 270 km. For the frequency difference $\delta v=100$ kHz, by applying a 10 GHz system and setting the matching time interval $T_c =1~ \mu $s, the secure transmission distance can still exceed 380 km. The corresponding loss is 62 dB.} In free space, if  the Micius satellite~\cite{chen2021integrated} is used as the intermediate station Charlie, the key distribution between two ground nodes with a distance of approximately 1000 km can be realized.  At an intercity distance of 300 km, the key rate is 0.15 Mbps, which is sufficient to perform a variety of tasks, including audio and video encryption. 	

\begin{figure}[t]
\centering
\includegraphics[width=8.6cm]{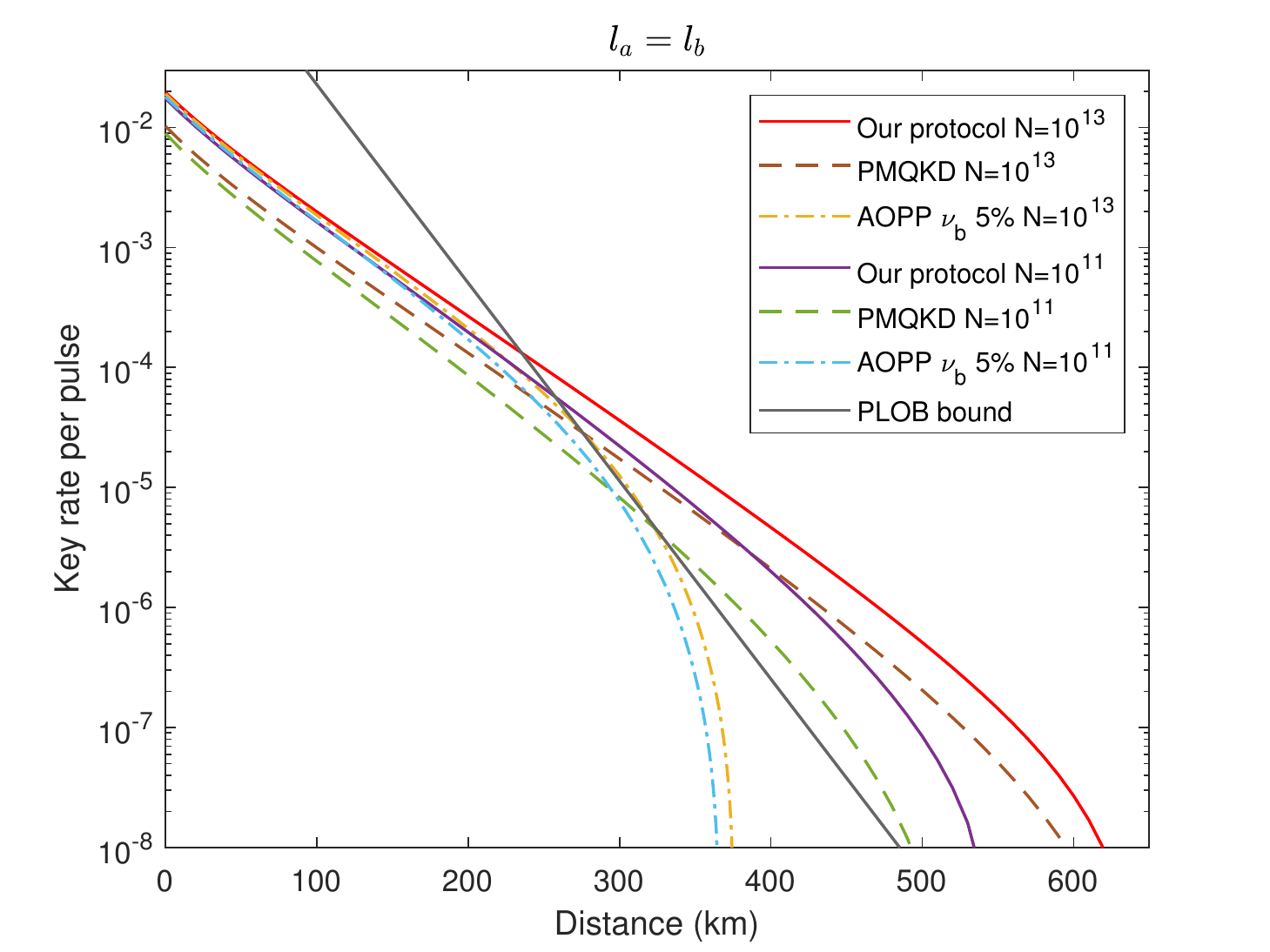}
\caption{Comparison of the secret key rates of the asynchronous-MDIQKD with arbitrary-time matching, PMQKD~\cite{zeng2020symmetry} and AOPP~\cite{xu2020sending} in symmetric channels. The numerical results here show 
	that our protocol has a notable advantage and is able to achieve a long transmission distance of 620 km.}\label{fig_keyrate_finite_sym}
\end{figure}	

{We remark that by circumventing the need for phase tracking, the asynchronous-MDIQKD protocol has a noteworthy advantage over TFQKD at intercity distance. Typically, the maximum counting rate of a commercially available SNSPD is approximately 2 MHz per channel. For TFQKD, using strong reference light to execute phase-tracking usually consumes a count rate of about 4 MHz per channel~\cite{wang2019beating} (sometimes 40 MHz peak count rate~\cite{liu2019experimental}), and dedicated high-performance detectors are needed. In Ref.~\cite{liu2019experimental}, the two-parallel-nanowire serial-connected configuration is developed to address the high count rate issue. In contrast, asynchronous-MDIQKD does not impose a strict count rate requirement on the detector, and all detector count rates are usable for quantum signals. Assuming the maximum counting rate is 5 MHz per channel, at 230 km, with a 4 GHz repetition rate, the count rate of the quantum signal will be approximately 4.4 MHz per channel, which can be used for key generation in our protocol. The key rate is 350 kbps when $\delta v = 10$ kHz and $T_c = 10~\mu $s. However, for TFQKD, the available count rate is only 1 MHz per channel, resulting in a key rate of approximately 20 kbps~\cite{liu2019experimental}. In this case, the key rate of our protocol is one order of magnitude higher than that of TFQKD.}

For a large time interval $T_c$, say $1$~s, the phase correlation between two time bins fades. With the same experimental complexity as TFQKD, that is, using phase locking and phase tracking, one can postmatch time bins with arbitrary time interval and achieve better performance.	Here, we simulate the key rates of  asynchronous-MDIQKD  with arbitrary time matching and compare it with  those of PMQKD (PMQKD)~\cite{ zeng2020symmetry} and SNSQKD with the help of actively AOPP. We set the total number of pulses as $N = 10^{11}$ and $10^{13}$,  the misalignment  in the $X$ basis as $\sigma=\pi/36 $, and the security bounds as $\varepsilon_{\rm{AMDI}}=\varepsilon_{\rm{AOPP}}=3.6\times 10^{-9}$ and $\varepsilon_{\rm{PM}}=O(10^{-9})$.  The detailed formulas for simulating our protocol are presented in Appendix~\ref{simu_pb}.

\begin{figure}[t]
\centering
\includegraphics[width=8.6cm]{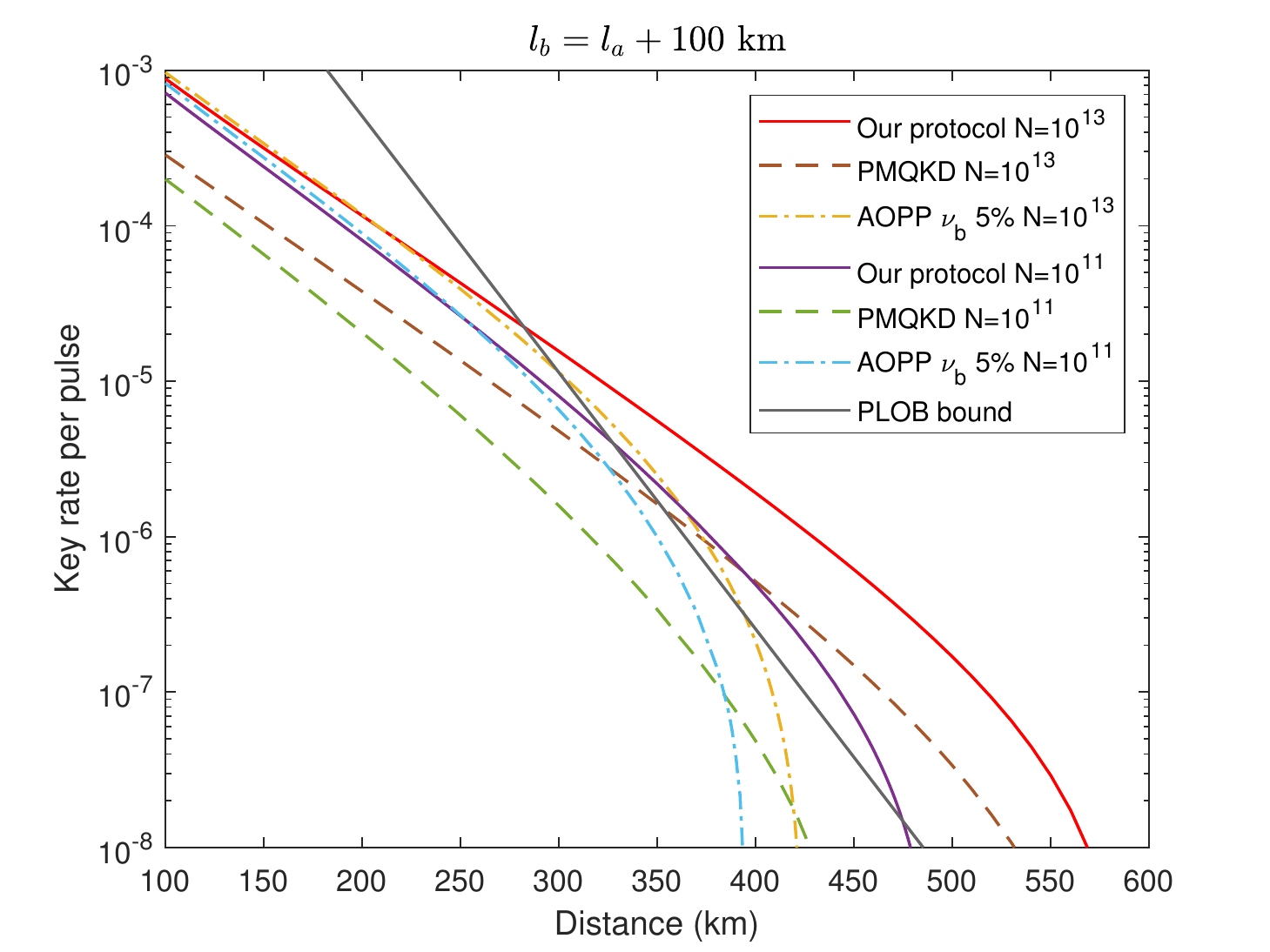}
\caption{Comparison of secret key rates of our protocol  with arbitrary time matching, PMQKD, and AOPP versus transmission distance in asymmetric channels. Our protocol exhibits  a decent performance in asymmetric channels. }\label{fig_keyrate_finite_asym}
\end{figure}

Because our protocol is an MDI-type protocol, the density matrix of the single-photon pair component is always identical in the X and Z bases for each user, regardless of the asymmetric source parameters chosen.  This makes it possible for a dynamic quantum network to add or delete new user nodes without considering the source parameters of existing users. In contrast, to guarantee that the density matrix of the two users' joint single-photon state in the X basis is the same as that in the Z basis for SNSQKD protocols, the transmission probability and intensity of the coherent state must follow strict mathematical constraint ~\cite{wang2018twin,hu2019sending}. However, this constraint is difficult to realize in practice, especially in networks where users are added and deleted over time, which greatly degrades their performance.  By exploiting the quantum coin concept~\cite{lo2007security,koashi2009simple}, a recent study provided a security proof for the SNSQKD protocol when the constraint is not satisfied~\cite{xie2021scalable}.
When comparing the key rates, we considered that the intensity of the coherent state in AOPP does not satisfy the mathematical constraint (which is often the case in practice) with a modulation deviation of decoy state $\nu_b$ to $5\%$, and the other parameters have no deviation. The key rates in symmetric channels are shown in Fig.~\ref{fig_keyrate_finite_sym}. The simulation results show that the key rate of asynchronous-MDIQKD  is always higher than that of PMQKD and AOPP. At 500 km, for $N = 10^{13}$, the secret key rate of our protocol is $150\%$ higher than that of PMQKD, and the transmission distance is 240 km longer than that of AOPP. For $N = 10^{11}$, our protocol transmits over a distance of more than 500 km. Fig.~\ref{fig_keyrate_finite_asym} shows the key rate in the asymmetric channels, where $l_b = l_a + 100$ km.	Notably, our protocol also performs well in asymmetric channels. At 500 km, for $N = 10^{13}$, the key rate of our protocol is $400\%$ higher than that of PMQKD, and the transmission distance is 150 km longer than that of AOPP. Similarly, when $ N = 10 ^{11}$, the transmission distance of the asynchronous-MDIQKD protocol is 50 km higher.

In summary, the asynchronous-MDIQKD protocol does not require complicated phase-locking and phase-tracking techniques, and it is resistant to imperfect intensity modulation. Therefore, an intercity quantum network is possible, where users can dynamically access and freely choose to perform asynchronous-MDIQKD with short time matching or arbitrary time matching.

\section{conclusion}
\label{sec_conclusion}	
In this work, we presented an asynchronous-MDIQKD protocol through time multiplexing. We realized $O(\sqrt{\eta})$ scaling of the key rate with asynchronous two-photon interference, thus surpassing the PLOB bound. By removing phase locking and phase tracking, our protocol greatly simplifies the hardware requirement with a small sacrifice in performance. 
When using the same experimental techniques as TFQKD, our protocol is secure against coherent attacks, and shows longer transmission and higher key rate than PMQKD and SNSQKD (AOPP), considering imperfect intensity modulation. Our work also suggests a practical method of overcoming the linear bound  of dual-rail protocols~\cite{das2021universal} without challenging technologies. In addition, our protocol can also exploit the six-state encoding~\cite{lo2001six} due to random phase modulation. Therefore, if applying single photon sources in the $Z$ basis, our protocol can be made secure up to higher error rate by establishing the non-trivial mutual information between the bit-flip and phase error patterns~\cite{lo2001six,yin2016security}, thereby achieving a higher key rate. This work exhibits remarkable superiority in intercity quantum network deployment for balancing performance and technical complexity. We believe the key contributions of this work will produce exciting opportunities for the widespread deployment of global quantum networks beyond quantum key distribution, ranging from quantum repeaters to quantum entanglement distribution. 

\par
\begin{center}
\textbf{ACKNOWLEDGMENTS} 
\end{center}

We gratefully acknowledge the support from the National Natural Science Foundation of China (No. 61801420), the Natural Science Foundation of Jiangsu Province (No. BK20211145), the Fundamental Research Funds for the Central Universities (No. 020414380182), the Key Research and Development Program of Nanjing Jiangbei New Aera (No. ZDYD20210101), the Key-Area Research and Development Program of Guangdong Province (No. 2020B0303040001), and the China Postdoctoral Science Foundation (No. 2021M691536).

{Note added--- During the peer review of our work, we became aware of a similar work by Zeng et al.~\cite{zeng2022quantum}, who consider a mode-pairing MDI-QKD scheme that matches the adjacent detection pulses to extract the key information. By assuming infinite decoy states, they calculated the key rate in the asymptotic regime, which can break the rate-loss bound. In this work, we use the three-intensity decoy-state method to calculate the key rate in the finite-size regime and show its ability to break the rate-loss bound.}

\appendix
\section{Simulation formulas}\label{app_simulation}

In this section, we calculate the  parameters in Eq.~\eqref{eq_keyrate_finite} to estimate the secret key rate.
In the following description, let $x^*$ be the expected value of $x$. We denote the number of  $\{k_a,~k_b\}$  as $x_{k_ak_b}$. We denote the number and  error number of events  $\{k_a^{i}k_a^{j},~k_b^{i}k_b^{j}\}$ after postmatching as $n_{k_a^{i}k_a^{j},~k_b^{i}k_b^{j}}$ and $m_{k_a^{i}k_a^{j},~k_b^{i}k_b^{j}}$, respectively.  For simplicity, we abbreviate  $k_a^ik_a^j,k_a^ik_a^j$ as $2k_a,2k_b$ when $k_a^i=k_a^j$ and $k_b^i=k_b^j$.

\textbf{1.~$\underline{s}_{11}^{z}$.}
$s_{11}^z$ corresponds to the number of successful detection events where Alice and Bob each emit a single photon  in different time bins in the $Z$ basis. We define $z_{10}$ ($z_{01}$) as the number of events in which Alice (Bob) emits a single photon and Bob (Alice) emits a vacuum state in $\{\mu_a,\mathbf{o}_b\}$ ($\{\mathbf{o}_a,\mu_b\}$) event. The lower bounds of their expected values are $\underline{z}_{10}^{*}= N p_{\mu_a} p_{\mathbf{o}_{b}} \mu_{a}e^{-\mu_a} \underline{y^*_{10}}$ and $
\underline{z}^*_{01} = N p_{\mathbf{o}_{a}} p_{\mu_b} \mu_{b} e^{-\mu_b} \underline{y^*_{01}} $, where the yields $\underline{y}_{10}^*$ and $\underline{y}_{01}^*$ are the corresponding yields. These can be estimated using the decoy-state method:
\begin{align}
\underline{y}_{01}^*\geq& \frac{\mu_b}{N(\mu_b\nu_b-\nu_b^2)} \left(\frac{e^{\nu_b}\underline{x}_{o_a\nu_b}^{*}}{p_{o_a}p_{\nu_b}}\nonumber\right.\\
&\left.-\frac{\nu_b^2}{\mu_b^2}  \frac{e^{\mu_b}\overline{x}_{\hat{\mathbf{o}}_{a}\mu_b}^{*}}{p_{\hat{\mathbf{o}}_{a}}p_{\mu_b}} - \frac{\mu_b^2-\nu_b^2}{\mu_b^2}\frac{\overline{x}_{o o}^{d*}}{p_{o_ao_b}^d}\right),\\
\underline{y}_{10}^*\geq&\frac{\mu_a}{N(\mu_a\nu_a-\nu_a^2)} \left( \frac{e^{\nu_a}\underline{x}_{\nu_ao_b}^{*}}{p_{\nu_a}p_{o_{b}}}\nonumber\right.\\
& \left.-\frac{\nu_a^2}{\mu_a^2} \frac{e^{\mu_a}\overline{x}_{\mu_a\hat{\mathbf{o}}_{b}}^{*}}{p_{\mu_  a}p_{\hat{\mathbf{o}}_{b}}}- \frac{\mu_a^2-\nu_a^2}{\mu_a^2}\frac{\overline{x}_{o o}^{d*}}{p_{o_ao_b}^d}\right),\label{eq_decoy_Y01}
\end{align}
where $x_{oo}^{d}=x_{\hat{\mathbf{o}}_{a}\hat{\mathbf{o}}_{b}}+x_{\hat{\mathbf{o}}_{a}\mathbf{o}_b}+x_{\mathbf{o}_{a}\hat{\mathbf{o}}_{b}}$ represents the number of events where at least one user chooses the declare-vacuum state and $p_{oo}^d=p_{\hat{\mathbf{o}}_{a}\hat{\mathbf{o}}_{b}}+p_{\hat{\mathbf{o}}_{a}\mathbf{o}_b}+p_{\mathbf{o}_{a}\hat{\mathbf{o}}_{b}}$ refers to the corresponding probability.  Thus, the lower bound of $s_{11}^{z*}$ can be given by
\begin{equation}
\begin{aligned}
	\underline{s}_{11}^{z*}=n_C^z\frac{\underline{z}_{10}^*}{x_{\mu_a\mathbf{o}_{b}}} \frac{\underline{z}_{01}^*}{x_{\mathbf{o}_{a}\mu_b}}= \frac{\underline{z}_{10}^*\underline{z}_{01}^*}{x_{\max}}.\\
\end{aligned}
\end{equation}
where  $x_{0}=x_{\mathbf{o}_{a}\mu_b}+x_{\mathbf{o}_{a}\mathbf{o}_{b}}$, $x_{1}  =x_{\mu_a\mathbf{o}_{b}}+x_{\mu_a\mu_b}$ and  $x_{\max}=\max\{x_{0},x_{1}\}$.

\textbf{2.~$\underline{s}_{0\mu_b}^{z}$.} $s_{0\mu_b}^z$ represents the number of events in the $Z$ basis, Alice emits a zero-photon state in the two matched time bins, and the total intensity of Bob's pulses is $\mu_b$. We define $z_{00}$ ($z_{0\mu_b}$) as the number of detection events where the state sent by Alice collapses to the vacuum state in the $\{\mu_a,\mathbf{o}_b\}$ ($\{\mu_a,\mu_b\}$) event. The lower bounds of the expected values are $\underline{z}_{00}^*={ p_{\mu_a} p_{\mathbf{o}_b}e^{-\mu_a}\underline{x}_{o o}^{d*}}/{p_{oo}^d}$ and $ ~\underline{z}_{0\mu_b}^*=  {p_{\mu_a} p_{\mu_b}e^{-\mu_a}\underline{x}_{\mathbf{o}_{a}\mu_b}^{*}}/{p_{\mathbf{o}_{a}}p_{\mu_b}}$, respectively. In this study, we employed the relation between the expected value $\underline{x}_{\mathbf{o}_{a}\mu_b}^{*}={ p_{\mathbf{o}_{a}} \underline{x}_{\hat{\mathbf{o}}_{a}\mu_b}^{*}}/{ p_{\hat{\mathbf{o}}_{a}}}$, and $~\underline{x}_{\mathbf{o}_{a}\mathbf{o}_b}^{*}={ p_{\mathbf{o}_{a}} p_{\mathbf{o}_b}\underline{x}_{o o}^{d*}}/{p_{oo}^d}$. The lower bound of $s_{0\mu_b}^{z*}$ can be written as
\begin{equation}
\begin{aligned}
	\underline{s}_{0\mu_b}^{z*}=& n_C^{z*} \frac{\underline{z}_{00}^*}{x_{\mu_a\mathbf{o}_{b}}^*} +n_E^{z*} \frac{\underline{z}_{0\mu_b}^*}{x_{\mu_a\mu_b}^*},\label{eq_asynchronous-MDIQKD_S0uz}
\end{aligned}
\end{equation}

\textbf{3.~$\underline{s}_{11}^{x}$.} 
The phase difference between Alice and Bob is defined as $\varphi= \theta_a- \theta_b +\phi$ and the corresponding number in the  $\{k_a,~k_b\}$ event as $x_{k_ak_b}^\varphi $. In the post-matching step, two time bins are matched if they have the same phase difference $\varphi$, and all $\{2\nu_a,~2\nu_b\}$ events can be grouped according to the phase difference $\varphi$. We denote the number of  $\{2\nu_a,~2\nu_b\}$ events with phase difference $\varphi$ as $n_{2\nu_a,2\nu_b}^\varphi=x_{\nu_a\nu_b}^\varphi/2$. Similar to the time-bin MDIQKD, the  expected yields of single-photon pairs in the $X$ and $Z$ bases satisfy the following relation:
\begin{equation}
\begin{aligned}
	Y_{11}^{x*}&=Y_{11}^{z*}=\frac{1}{4}(y_{01}^*y_{10}^*+
	y_{10}^*y_{01}^*+y_{00}^*y_{11}^*+y_{11}^*y_{00}^*)
	\\
	&\ge \frac{1}{2}y_{10}^*y_{01}^*.	
\end{aligned}
\end{equation}
Suppose the global phase difference  $\varphi$  is a randomly and uniformly distributed value. The expected number of single-photon pairs can be given by
\begin{equation}
\begin{aligned}
	s_{11}^{x*}&=\frac{1}{2\pi}\int_{0}^{2\pi}n_{2\nu_a,2\nu_b}^\varphi\times \frac{4\nu_a\nu_be^{-2\nu_a-2\nu_b }}{q_{ \nu_a\nu_b }^\varphi q_{\nu_a \nu_b }^\varphi} y_{11}^{x*}d\varphi\\
	&\ge\frac{1}{2\pi}\int_{0}^{2\pi}n_{2\nu_a,2\nu_b}^\varphi\times \frac{\nu_a\nu_be^{-2\nu_a-2\nu_b }}{q_{ \nu_a\nu_b }^\varphi q_{\nu_a \nu_b }^\varphi} 2\underline{y}_{10}^*\underline{y}_{01}^*d\varphi\\
	&=N p_{\nu_a }p_{\nu_b }\nu_a \nu_b e^{-2(\nu_a +\nu_b )}\underline{y}_{ 10}^{*}\underline{y}_{ 01}^{*}\int_{0}^{2\pi}\frac{1}{2\pi q_{\nu_a \nu_b}^\varphi}d\varphi,
	\label{eq_AMIDQKD_s11x2}	
\end{aligned}
\end{equation}
where the coefficient 4 on the first line of the formula corresponds to the four modes of single-photon pairs in the $X$ basis, $q_{\nu_a \nu_b }^\varphi$ is the gain when Alice chooses intensity $\nu_a $, and Bob chooses intensity $\nu_b$ with phase difference $\varphi$ and $n_{2\nu_a,2\nu_b}^\varphi$$=$$N p_{\nu_a}p_{\nu_b}q_{\nu_a \nu_b }^\varphi/2$. We define $q_{k_a k_b }$$=$$1/(2\pi)\int_{0}^{2\pi} q_{k_a k_b }^\varphi d\varphi$ as the average gain, given that Alice chooses intensity $k_a $, and Bob chooses intensity $k_b$. When $\nu_a$ and $\nu_b\approx0$, there is an approximation relation
$\int_{0}^{2\pi} 1/ (2\pi q_{\nu_a\nu_b}^\varphi)d\varphi\approx 1/ q_{\nu_a\nu_b}$. 

In the discrete case, the phase difference $\varphi$ is divided into $M$ slices $\{\delta_m\}$ for $1\le m\le M$, where $m$ is an integer, where $\delta_m=[2\pi (m-1)/M,2\pi m/M)$. The expected number of single-photon pairs is given by 
\begin{equation}
\begin{aligned}
	\underline{s}_{11}^{x*}&\ge\sum_{m=1}^M n_{2\nu_a,2\nu_b}^m\times 2\frac{\nu_ae^{-\nu_a-\nu_b }\underline{ y}_{ 10}^{*}}{q_{ \nu_a\nu_b }^m} \frac{\nu_b e^{-\nu_a -\nu_b }\underline{ y}_{01}^{*} }{ q_{\nu_a \nu_b }^m},\\
\end{aligned}
\end{equation}
where $n_{2\nu_a,2\nu_b}^m$ is the number of $\{2\nu_a,2\nu_b\}$ events with phase difference $\varphi$ falling into slice $\delta_m$. $q_{\nu_a \nu_b }^m$ is the corresponding gain.

\textbf{4.~$\overline{e}_{11}^{x}$.} 
For single-photon pairs, the expected value of the phase error rate in the $Z$ basis equals the expected value of the bit error rate in the $X$ basis, and the error rate $\overline{e}_{11}^{x} = {\overline{t}_{11}^{x}}/{\underline{s}_{11}^x}$. Therefore, we first calculate the number of errors of the single-photon pairs in the $X$ basis ${t_{11}^x}$. The upper bound of ${t_{11}^x}$  can be expressed as 
\begin{equation}
\begin{aligned}
	\overline{t_{11}^x}\leq& m_{2\nu_a,2\nu_b} - (\underline{m_{0, 2\nu_b}+m_{2\nu_a, 0}}) +\overline{m}_{0 ,0}, 	
\end{aligned}\label{eq_asynchronous-MDIQKD_t11}
\end{equation}
where  $m_{0, 2\nu_b}$ ($m_{2\nu_a, 0}$) is the error count when the state sent by Alice (Bob) collapses to the vacuum state in events $\{2\nu_a,2\nu_b\}$, and $m_{0,0}$ corresponds to the event where the states sent by Alice and Bob both collapse to vacuum states in events $\{2\nu_a,2\nu_b\}$. 
The expected error counts $m_{0, 2\nu_b}^*$ and $m_{2\nu_a, 0}^*$ can be expressed as follows:
\begin{equation}
\begin{aligned}
	m_{0, 2\nu_b}^*&=e_o\frac{1}{2\pi}\int_{0}^{2\pi}n_{2\nu_a,2\nu_b}^\varphi\frac{e^{-\nu_a}q_{0\nu_b}^{*}}{q_{ \nu_a\nu_b}^\varphi}\frac{e^{-\nu_a}q_{0\nu_b}^{ *}}{q_{\nu_a\nu_b}^{\varphi}}d\varphi\\
	&= e_oN p_{\nu_a}p_{\nu_b}e^{-2\nu_a}q_{0\nu_b}^{ *2}\frac{1}{4\pi}\int_{0}^{2\pi}\frac{1}{q_{\nu_a\nu_b}^\varphi}d\varphi,\\
	m_{2\nu_a, 0}^*&=e_o\frac{1}{2\pi}\int_{0}^{2\pi}n_{2\nu_a,2\nu_b}^\varphi\frac{e^{-\nu_b}q_{\nu_a0}^{*}}{q_{ \nu_a\nu_b}^\varphi}\frac{e^{-\nu_b}q_{\nu_a0}^{ *}}{q_{\nu_a\nu_b}^{\varphi}}d\varphi\\
	&= e_oN p_{\nu_a}p_{\nu_b}e^{-2\nu_b}q_{\nu_a0}^{ *2}\frac{1}{4\pi}\int_{0}^{2\pi}\frac{1}{q_{\nu_a\nu_b}^\varphi}d\varphi,\\
\end{aligned}
\end{equation}
respectively, where $e_o=1/2$ is the error rate of the background noise. 

In the symmetric case, $\nu_a = \nu_b$, $p_{o_a}=p_{o_b}$, and $~p_{\nu_a}=p_{\nu_b}$. In this case, we have $q_{\nu_ao}^*=q_{o\nu_b}^*=(x_{o_a\nu_b}+x_{\nu_ao_b})^*/(2 Np _{o_a}p_{\nu_b})$.
In the asymmetric case, $q_{\nu_ao}^*=x_{\nu_ao_b}^*/(Np_{\nu_a}p_{o_b})$ and $q_{o\nu_b}^*=x_{o_a\nu_b}^*/(Np_{o_a}p_{\nu_b})$. 
Then, the lower bound of the observed value of $m_{0, 2\nu_b}+m_{2\nu_a, 0}$ can be written as
\begin{equation}
\begin{aligned}\label{eq_m02v_calcu}
	\underline{m_{0 ,2\nu_b}+m_{2\nu_a 0}}=&
	\varphi^L(\underline{m}_{0 ,2\nu_b}^*+\underline{m}_{2\nu_a ,0}^*, \epsilon).
\end{aligned}
\end{equation}
where  $\varphi^L(x)$ is the lower bounds when using Chernoff bound to
estimate the real values according to the expected values  and is defined in Eq.~\ref{chernoff1}.	
The expected value of ${m}_{0, 0}$ can be given by
\begin{equation}
\begin{aligned}	
	{m}_{0, 0}^*= &e_o\frac{1}{2\pi}\int_{0}^{2\pi}n_{2\nu_a,2\nu_b}^\varphi\frac{e^{-\nu_a-\nu_b}q_{00}^{*}}{q_{ \nu_a\nu_b}^\varphi}\frac{e^{-\nu_a-\nu_b}q_{00}^{ *}}{q_{\nu_a\nu_b}^{\varphi}}d\varphi\\
	&= e_oN p_{\nu_a}p_{\nu_b}e^{-2(\nu_a+\nu_b)}q_{00}^{ *2}\frac{1}{4\pi}\int_{0}^{2\pi}\frac{1}{q_{\nu_a\nu_b}^\varphi}d\varphi.\\
\end{aligned}
\end{equation}
The upper bound of $q_{00}^*$ can be obtained from  $\overline{q}_{00}^*=\overline{m}_{oo}^{d*}/(Np_{oo}^d)$.
The upper bound of $m_{0, 0}$ can be obtained by  $\overline{m}_{0, 0}=\varphi^U(\overline{m}_{0, 0}^*, \epsilon)$, where  $\varphi^U(x)$ is the upper bound while using the Chernoff bound to estimate the observed values according to the expected values and is defined in Eq.~\ref{chernoff2}.	

\textbf{5.~$\overline{\phi}_{11}^{z}$.}
Finally,  for a failure probability $\varepsilon$, the upper bound of the phase error rate $\overline{\phi}_{11}^{z}$ can be obtained  
by using random sampling without replacement in~Eq. \eqref{Randomswr}
\begin{equation}
\begin{aligned}
	\overline{\phi}_{11}^{z}\leq& \overline{e}_{11}^x	+\gamma \left(\underline{s}_{11}^z,\underline{s}_{11}^x,\overline{e}_{11}^x,\varepsilon\right).\\
\end{aligned}\label{eq_asynchronous-MDIQKD_phi11z}
\end{equation}

\section{Simulation details}\label{simu_pb}

\subsection{Asynchronous-MDIQKD protocol for arbitrary-time matching}\label{simuarbitrary}

Similar to the time-bin encoding MDIQKD, the valid events after postmatching in the $Z$ basis can be divided into  correct events $\{\mu_a \mathbf{o}_{a},~ \mathbf{o}_{b}\mu_b\}$, $\{\mathbf{o}_{a}\mu_a,~\mu_b \mathbf{o}_{b}\}$,  and incorrect events $\{\mu_a\mathbf{o}_{a},~\mu_b\mathbf{o}_{b}\}$, $\{\mathbf{o}_{a}\mu_a,~ \mathbf{o}_{b}\mu_b\}$. The corresponding numbers are denoted as
$n_C^z$ and $n_E^z$, respectively, which can be written as
\begin{align}
n_C^z = x_{\min}\frac{x_{\mathbf{o}_{a} \mu_b}}{x_{0}}\frac{x_{\mu_a\mathbf{o}_{b}}}{x_{1}} =  \frac{x_{\mathbf{o}_{a} \mu_b}x_{\mu_a \mathbf{o}_{b}}}{x_{\max}}\nonumber,
\end{align} 
and 
\begin{align}
n_E^z = x_{\min}\frac{x_{\mathbf{o}_{a} \mathbf{o}_{b}}}{x_{0}}\frac{x_{\mu_a\mu_b}}{x_{1}}= \frac{x_{\mathbf{o}_{a} \mathbf{o}_{b}} x_{\mu_a \mu_b}}{x_{\max}}\nonumber,
\end{align}  
where $x_{\min}=\min\{x_{0},x_{1}\}$.   
The overall number of events in the $Z$ basis is $n^z = n^z_C+n^z_E$ and the bit error rate in the $Z$ basis is $E^z={n^z_E}/{n^z}$.

In the $X$ basis, the data are composed of events $\{2\nu_a,~2\nu_b\}$, $\{2o_a,~2\nu_b\}$, $\{2\nu_a,~2o_b\}$,
$\{2\hat{\mathbf{o}}_a,2o_b\}$, and $\{2\mathbf{o}_a,~2\hat{\mathbf{o}}_b\}$. Without loss of generality, we consider the case in which all matched events satisfy $\theta_a^i -\theta_a^j - (\theta_b^i - \theta_b^j)+(\phi^i - \phi^j) = 0$.
In this case,  when $r_a^i \oplus r_a^j \oplus r_b^i \oplus r_b^j=0$ (1), the $\{\nu_a^i\nu_a^j,~\nu_b^i\nu_b^j\}$ event is considered to be an error event when different detectors  (the same detector)  click at time bins $i$ and $j$. 

When Alice and Bob send intensities $k_a$ and $k_b$ with  phase difference $\varphi$, the gain corresponding to only one detector click is
\begin{equation}
\begin{aligned}
	q_{k_ak_b}^{L\varphi}=&y_{k_ak_b}\left[e^{\omega_{k_ak_b}\cos\varphi}-y_{k_ak_b}\right],\\
	q_{k_ak_b}^{R\varphi}=&y_{k_ak_b}\left[e^{-\omega_{k_ak_b}\cos\varphi}-y_{k_ak_b}\right].\\
\end{aligned}
\end{equation}
where $y_{k_ak_b}:=e^{\frac{-(\eta_ak_a+\eta_bk_b)}{2}}(1-p_d)$, $\omega_{k_ak_b}$$:=$$\sqrt{\eta_ak_a\eta_bk_b}$, $\eta_a=\eta_d10^{-\alpha l_a/10}$ and $\eta_b=\eta_d10^{-\alpha l_b/10}$.   The overall gain can be given by $q_{k_ak_b}= 1/2\pi\int_{0}^{2\pi}q_{k_ak_b}^{\varphi}d\varphi=1/2\pi\int_{0}^{2\pi}(q_{k_ak_b}^{L\varphi}+q_{k_ak_b}^{R\varphi})d\varphi= 2y_{k_ak_b}[I_0(\omega_{k_ak_b})-y_{k_ak_b}]$, where $I_0(x)$ represents the zero-order modified Bessel function of the first kind. The total number of $\{k_a,k_b\}$ is $x_{k_ak_b}=Np_{k_a}p_{k_b}q_{k_ak_b}$.

The overall error count in  the $X$ basis can be given as 
\begin{equation}
\begin{aligned}
	m_{2\nu_a,2\nu_b}&=\frac{1}{2\pi}\int_{0}^{2\pi}n_{2\nu_a2\nu_b}^\varphi\left[\frac{q_{\nu_a\nu_b}^{L\varphi}q_{\nu_a\nu_b}^{R({\varphi+\sigma})}+q_{\nu_a\nu_b}^{R\varphi}q_{\nu_a\nu_b}^{L({\varphi+\sigma})}}{q_{ \nu_a\nu_b}^\varphi q_{ \nu_a\nu_b}^{\varphi+\sigma}}\right]d\varphi\\
	&=N p_{\nu_a}p_{\nu_b}\frac{1}{4\pi}\int_{0}^{2\pi}\frac{q_{\nu_a\nu_b}^{L\varphi}q_{\nu_a\nu_b}^{R({\varphi+\sigma})}+q_{\nu_a\nu_b}^{R\varphi}q_{\nu_a\nu_b}^{L({\varphi+\sigma})}}{q_{ \nu_a\nu_b}^{\varphi+\sigma}}d\varphi.
\end{aligned}
\end{equation}
where $\sigma$ refers to the angle of misalignment in the $X$ basis.

\subsection{Asynchronous-MDIQKD protocol for short-term matching}\label{simple_simu}
The total number of time bins per $T_c$ is $N_{T_c} =T_cF$. 
For each detection event, the probability that it belongs to $\mathit{case~1}$ is $p_{c_1}=1-(1-\bar{p})^{2N_{T_c}-2}$ and  the probability of belonging to $\mathit{case~2}$ is $p_{c_2}=(1-\bar{p})^{2N_{T_c}-2}$, where $\bar{p}=\sum_{k_a,k_b}p_{k_a}p_{k_b}q_{k_ak_b}$ is the average detection probability, and $2N_{T_c}-2$ is the total number of time bins neighboring the given detection event. The mean number of $\mathit{case~2}$ events is $\bar{N}^{c_2}=Np_{c_2}\bar{p}$.

For simplicity, we divide the post-matching window into $d$ windows with a time length of $T_c$, where $d=N/N_{T_c}$.
In the $Z$ basis, given that Alice sends $k_a$ and Bob sends $k_b$, the detection count in the $i$-th window and the total detection count are $x_{k_ak_b}^i=N_{T_c}p_{c_1}p_{k_a}p_{k_b}q_{k_ak_b}$  and $x_{k_ak_b}=Np_{c_1}p_{k_a}p_{k_b}q_{k_ak_b}$, respectively.  After postmatching, the number of correct and incorrect events in the $Z$ basis  is
\begin{equation}
\begin{aligned}
	n^{z}_C =\sum^d_{i=1} x^{i}_{\min} \frac{x^{i}_{\mathbf{o}_{a} \mu_b}}{x^{i}_{0}}\frac{x^{i}_{\mu_a\mathbf{o}_b}}{x^{i}_{1}}=\sum^d_{i=1} \frac{x^{i}_{\mu_a\mathbf{o}_b}x^{i}_{\mathbf{o}_{a} \mu_b}}{x^{i}_{\max}}\nonumber,\\
	n^{z}_E =\sum^d_{i=1}  x^{i}_{\min} \frac{x^{i}_{\mathbf{o}_{a} \mathbf{o}_b}}{x_{0}^{c_1}}\frac{x_{\mu_a\mu_b}^{i}}{x_{1}^{i}}=\sum^d_{i=1} \frac{x_{\mathbf{o}_{a} \mathbf{o}_b}^{i} x_{\mu_a \mu_b}^{i}}{x^{i}_{\max}},
\end{aligned}
\end{equation}
respectively, where $x^{i}_{0}=x^{i}_{\mathbf{o}_{a}\mu_b}+x^{i}_{\mathbf{o}_{a}\mathbf{o}_b}$, $x^{i}_{1}=x^{i}_{\mu_a\mathbf{o}_b}+x^{i}_{\mu_a\mu_b}$, $x^{i}_{\min}=\min\{x^{i}_{0},x^{i}_{1}\}$, $x^{i}_{\max}=\max\{x^{i}_{0},x^{i}_{1}\}$. 
The overall number of events in the $Z$ basis is $n^{z} =  n^{z}_C+n^{z}_E$, and the bit error rate in the $Z$ basis is $E^z={n^z_E}/{n^z}$. 	

In our simulation, we set $M=16$. Assuming that $x_{\nu_a\nu_b}^m$, the detection count of $\{\nu_a,~\nu_b\}$ events in slice $\delta_m$ follows the Poisson distribution  $P_r(x_{\nu_a\nu_b}^m=j)= \frac{\lambda^j}{j!}e^{-\lambda}$, where $\lambda=\bar{x}_{\nu_a\nu_b}^{i}$ is the mean value of $x_{\nu_a\nu_b}^m$. In the postmatching step, if $x_{\nu_a\nu_b}^m$ is even, all $\{\nu_a,~\nu_b\}$ events within the $m$-th $T_c$ will be utilized. If $x_{\nu_a\nu_b}^m$ is odd, there will be a redundant $\{\nu_a,~\nu_b\}$ event to be aborted. Therefore, the mean number of $\{2\nu_a,~2\nu_b\}$ events per $T_c$  is
\begin{equation}
\begin{aligned}			n_{2\nu_a,2\nu_b}^{i}&=\frac{2}{M}\left[\sum_{k}^{\left \lfloor \frac{N_{T_c}-1}{2} \right \rfloor}
	kP_r(2k)+kP_r(2k+1)\right]\\		&=\frac{1}{M}\left[\lambda-\frac{1-e^{-2\lambda}}{2}\right].
\end{aligned}
\end{equation}

The overall error count in  the $X$ basis can be given as 
\begin{equation}
\begin{aligned}
	m_{2\nu_a,2\nu_b}^{i}&=n_{2\nu_a,2\nu_b}^{i}\sum_{m=0}^{\frac{M}{2}-1}\left[\frac{2}{M}\frac{q_{\nu_a\nu_b}^{L\varphi_m}q_{\nu_a\nu_b}^{R(\varphi_m+\sigma)}+q_{\nu_a\nu_b}^{R\varphi_m}q_{\nu_a\nu_b}^{L({\varphi_m+\sigma})}}{q_{ \nu_a\nu_b}^{\varphi_m }q_{ \nu_a\nu_b}^{\varphi_m+\sigma}}\right],\\
\end{aligned}
\end{equation}
where $\varphi_m=2\pi m/M$.

\section{Statistical fluctuation analysis}\label{statistical}
In this Appendix, we introduce the statistical fluctuation analysis method~\cite{yin2020tight} used in the simulation. 

{\it{~Chernoff bound.}}
Let $x^{*}$ be the  expected value of $x$.  For a given expected value $x^{*}$, the Chernoff bound can be used to obtain the upper and lower bounds of the observed value.
\begin{equation}
\begin{aligned}\label{chernoff1}
	\overline{x}&=\varphi^U(x^{*})=x^{*}+\frac{\beta}{2}+\sqrt{2\beta x^{*}+\frac{\beta^{2}}{4}},
\end{aligned}
\end{equation}
and
\begin{equation}
\begin{aligned}\label{chernoff2}
	\underline{x}&=\varphi^L(x^{*})=x^{*}-\sqrt{2\beta x^{*}},
\end{aligned}
\end{equation}
where $\beta=\ln{\epsilon^{-1}}$.

{\it{~Variant of Chernoff bound.}}
For a given observed value $x$ and failure probability $\varepsilon$, the upper and lower bounds of $x^{*}$ can be acquired by the variant of the Chernoff bound
\begin{equation}
\begin{aligned}\label{varchernoff1}
	\overline{x}^{*}&=x+\beta+\sqrt{2\beta x+\beta^{2}},\\
\end{aligned}
\end{equation}
and
\begin{equation}
\begin{aligned}\label{varchernoff2}
	\underline{x}^{*}&=\max\left\{x-\frac{\beta}{2}-\sqrt{2\beta x+\frac{\beta^{2}}{4}},~0\right\}.\\
\end{aligned}
\end{equation}

{\it{Random sampling without replacement.}}
Let $X_{n+k} := \{x_1, x_2, \dots, x_{n+k}\}$ be a string of binary bits of size $n + k$,
where the number of bits is unknown. Let $X_k$ be a random sample (without replacement) bit string with
$k$ is picked from $X_{n+k}$ . Let $\lambda$ be the probability of a bit value 1 observed in $X_k$. Let $X_n$ be the remaining bit string, where the probability of bit value 1 observed in $X_n$ is $\chi$. The upper bound of 
$\chi$ can be expressed as
\begin{equation}
\begin{aligned}
	\overline{\chi}\leq&\lambda	+\gamma^{U} (n,k,\lambda,\epsilon),\\
\end{aligned}\label{Randomswr}
\end{equation}
where
\begin{equation}
\gamma^{U}(n,k,\lambda,\epsilon)=\frac{\frac{(1-2\lambda)AG}{n+k}+
	\sqrt{\frac{A^2G^2}{(n+k)^2}+4\lambda(1-\lambda)G}}{2+2\frac{A^2G}{(n+k)^2}},
\end{equation}
with $A=\max\{n,k\}$ and $G=\frac{n+k}{nk}\ln{\frac{n+k}{2\pi nk\lambda(1-\lambda)\epsilon^{2}}}$.

%\bibliography{bib}


\end{document}